\documentclass[12pt]{article}
\pdfoutput=1

\usepackage{jheppub}
\usepackage{amssymb,amsfonts,amsmath,amsthm}
\usepackage{mathrsfs}
\usepackage{bbold}
\usepackage{tikz}

\newcommand{\be}{\begin{equation}}
\newcommand{\ee}{\end{equation}}
\newcommand{\bea}{\begin{eqnarray}}
\newcommand{\eea}{\end{eqnarray}}

\newcommand{\CA}{\mathcal{A}}
\newcommand{\CB}{\mathcal{B}}
\newcommand{\CC}{\mathcal{C}}
\newcommand{\CD}{\mathcal{D}}
\newcommand{\CF}{\mathcal{F}}

\newcommand{\CN}{\mathcal{N}}
\newcommand{\CM}{\mathcal{M}}
\newcommand{\CO}{\mathcal{O}}
\newcommand{\CP}{\mathcal{P}}

\newcommand{\CT}{\mathcal{T}}

\newcommand{\CV}{\mathcal{V}}
\newcommand{\CW}{\mathcal{W}}

\newcommand{\lr}{\left (}
\newcommand{\rr}{\right )}
\newcommand{\ls}{\left [}
\newcommand{\rs}{\right ]}
\newcommand{\lc}{\left \{}
\newcommand{\rc}{\right \}}

\newcommand\qt\tau

\newcommand{\p}{\partial}

\renewcommand{\tilde}[1]{\widetilde{#1}}
\newcommand{\tr}{\text{tr}}

\makeatletter
\renewcommand{\@seccntformat}[1]{\csname the#1\endcsname.\,\,}
\makeatother
\allowdisplaybreaks

\let \savenumberline \numberline
\def \numberline#1{\savenumberline{#1.}}

\makeatletter
\def\@fpheader{\relax}
\makeatother

\def\bea{\begin{eqnarray}}
\def\eea{\end{eqnarray}}

\title{\ \vspace{1.6cm} \\
\scalebox{0.91}{Nonrelativistic Open String and Yang-Mills Theory}}
\author{Jaume Gomis${}^a$, Ziqi Yan${}^{a, b}$, and Matthew Yu${}^a$}
\emailAdd{jgomis@perimeterinstitute.ca}
\emailAdd{ziqi.yan@su.se}
\emailAdd{myu@perimeterinstitute.ca}
\affiliation{
${}^a$ Perimeter Institute for Theoretical Physics\\
        31 Caroline St N, Waterloo, ON N2L 2Y5, Canada \medskip\\
${}^b$ Nordita, KTH Royal Institute of Technology and Stockholm University\\
Roslagstullsbacken 23, SE-106 91 Stockholm, Sweden}
\abstract{The  classical and quantum worldsheet theory describing nonrelativistic open string theory   in an arbitrary nonrelativistic open and closed string background is constructed.  We show that the low energy dynamics of open strings ending   on $n$  coincident  D-branes in flat spacetime is described by a Galilean invariant  $U(n)$ Yang-Mills theory. We also study nonrelativistic open string  excitations  with winding number and demonstrate  that their dynamics  can be  encoded into  a local gauge theory in one higher dimension. By demanding conformal invariance of the boundary couplings, the nonlinear equations of motion that govern the consistent open string backgrounds coupled to an arbitrary closed background (described by a string Newton-Cartan geometry, Kalb-Ramond, and dilaton  field) are derived    and shown to emerge from a Galilean invariant Dirac-Born-Infeld type action.
}

\begin{document}

\maketitle
\vfill\eject
\section{Introduction}

Relativistic string theory is governed by the most general classically marginal 2d relativistic quantum field theory (QFT) with Poincar\'e symmetry realized as a global symmetry acting on the worldsheet fields: the nonlinear sigma model \cite{Friedan:1980jm}. In the presence of a boundary, additional marginal vertex operators supported on the boundary can be turned on, which describe a condensate of open strings ending on    D-branes \cite{Leigh:1989jq}. The coupling constants of the 2d QFT emerge as spacetime fields, and their properties are determined by the structure of the 2d QFT.  In particular, relativistic strings   propagate on a   Lorentzian Riemannian manifold, and in the presence of $n$ D-branes,  on  a background $U(n)$ vector bundle.\footnote{Not all   string theory backgrounds have a straightforward    geometric interpretation (e.g.   asymmetric orbifolds). They are nonetheless consistent classical string vacua as long as the 2d QFT is a CFT.} The celebrated field equations describing the propagation of massless particles of various helicities emerge from relativistic string theory at low energies by demanding quantum consistency of the 2d QFT on the string worldsheet, that is,  by imposing quantum conformal invariance  of the bulk and boundary couplings.  These    include  the Einstein, Rarita-Schwinger, Yang-Mills and Dirac equations. For a single D-brane, conformal invariance leads to a nonlinear theory for the curvature of the $U(1)$ connection: the Dirac-Born-Infeld (DBI) action      \cite{Leigh:1989jq,Callan:1986bc}.

Nonrelativistic string theory in flat spacetime was formulated in  \cite{Gomis:2000bd} as a 2d relativistic QFT on the worldsheet with a nonrelativistic global symmetry, known as string Newton-Cartan symmetry, acting on the string world sheet fields.\footnote{The nonrelativistic spectrum was first obtained by taking a limit of relativistic string theory in \cite{Klebanov:2000pp} (see also \cite{Danielsson:2000gi}).}  Realizing the string Newton-Cartan symmetry requires introducing   additional worldsheet fields, which  give  nonrelativistic string theory  some of its most salient features \cite{Gomis:2000bd}.  Nonrelativistic string theory is
governed by the most general  classically marginal 2d relativistic QFT with string Newton-Cartan symmetry acting on the worldsheet fields. The target space geometry induced by the 2d QFT couplings is the string Newton-Cartan geometry \cite{Bergshoeff:2018yvt} (see also \cite{Gomis:2005pg,Brugues:2004an,Andringa:2012uz,Brugues:2006yd,Bergshoeff:2019pij}),\footnote{For nonrelativistic geometries  from null reduction of relativistic string theory see  \cite{Harmark:2019upf}  and \S\ref{sec:closed}.} which is to nonrelativistic string theory what Lorentzian Riemannian geometry is to relativistic string theory. The equations of motion that determine the closed string backgrounds in which nonrelativistic strings can consistently propagate were derived in 
\cite{Gomis:2019zyu}. These are to nonrelativistic string theory what the (super)gravity equations of motion are to relativistic closed string theory.

In this paper, we study nonrelativistic open string theory in open and closed  string backgrounds.\footnote{Nonrelativistic open string theory corresponds to choosing a Dirichlet boundary condition along the longitudinal spatial direction \cite{Danielsson:2000mu} while a Neumann boundary condition \cite{Gomis:2000bd} leads to noncommutative open string theory \cite{Seiberg:2000ms,Gopakumar:2000na}. See \S\ref{sec:flat}.}  We determine the open string background fields in nonrelativistic open string theory by studying the space of open string vertex operators,   and derive the backgrounds in which   nonrelativistic open string theory can be consistently defined quantum mechanically. Our analysis leads to  interesting gauge theories with nonrelativistic symmetry living on the   D-branes on which nonrelativistic open strings end.

We derive from a worldsheet analysis  a nonlinear  $U(n)$ Yang-Mills action  with Galilean symmetry\footnote{More precisely, Bargmann symmetry. See \S\ref{sec:flat}.} in flat spacetime,
\be 
    S_\text{YM} = \frac{1}{g^2_\text{YM}} \int dX^0 \, dX^{A'} \, \tr \Bigl( \tfrac{1}{2} \, D_0 N\,D_0 N - E_{A'}D_{\!A'}N - \tfrac{1}{4}F_{A'B'}F_{A'B'} \Bigr),
    \label{formintro}
\ee
where $E_{A'}$ and $F_{A'B'}$ are gauge covariant electric and magnetic fields and $N$ is a scalar in the adjoint representation. The   theory for $n=1$ living on a single D-brane is quadratic and reduces to  Galilean Electrodynamics \cite{Santos:2004pq,Bergshoeff:2015sic, Festuccia:2016caf}. The gauge theory \eqref{formintro} describes the most general open string background  with vanishing winding number to lowest order in the $\alpha'$ expansion.
 
We   determine the spacetime effective gauge theory describing open string fields with nontrivial winding number. In spite that winding   introduces inherent nonlocalities in string theory, we show that the nonlocality can be tamed by  introducing an additional spacetime dimension on the D-brane, which has the physical interpretation as the coordinate conjugate to winding number. We derive the equations of motion describing the   background fields coupling to wound vertex operators and show that the effective action we derive in one higher dimension beautifully reproduces the spectrum and mass shell condition of the corresponding excitations of nonrelativistic open string theory.

Finally, we couple nonrelativistic string theory to an arbitrary closed and open string background. By imposing quantum mechanical conformal invariance on the open string couplings, we find a nonlinear system of equations of motion describing the open string fields. We show that these equations of motion can be derived from a nonlinear local field theory which has Galilean symmetry and which we dub Galilean DBI. 
 
The plan for the rest of the paper is as follows. In the first half of \S\ref{sec:flat}, we introduce nonrelativistic string theory in flat spacetime and classify the open string vertex operators. In the second half of \S\ref{sec:flat}, we determine the nonrelativistic gauge theories living  on D-branes by requiring that the open string vertex operators   be conformally invariant at the quantum level. In \S\ref{sec:curved}, we study nonrelativistic string theory in an arbitrary closed and open string background and derive the one-loop beta-functions for the boundary coupling constants in the corresponding Dirichlet sigma model. In \S\ref{sec:NRDBI}, we put forward the Galilean DBI action whose equations of motion reproduce the vanishing beta-functions. In \S\ref{sec:conclusions}, we conclude our paper.

\section{Flat Spacetime} \label{sec:flat}

\subsection{Nonrelativistic Open String Theory} \label{sec:nonrelost}

The  worldsheet action of nonrelativistic string theory  with  flat target spacetime in  conformal gauge is   \cite{Gomis:2000bd} 
\be \label{eq:Snonrel}
    	S_\text{flat} = \frac{1}{4 \pi \alpha '} \int_{\Sigma} d^2 z \bigl( 2 \, \p_z X^{A'} \p_{\overline{z}} X^{A'}  + \lambda \, \p_{\overline{z}} X + \overline{\lambda} \, \p_z \overline{X} \bigr)\,,
\ee
where
\be
    	X = X^0 + X^1\,,
        		\qquad
    	\overline{X} = X^0 - X^1\,,
\ee
and 
\be
	z = \sigma + i \tau\,,
		\qquad
	\overline{z} = \sigma - i \tau\,.
\ee
It follows that
\be
    \p_z = \tfrac{1}{2} \bigl( \p_\sigma - i \, \p_\tau \bigr)\,,
        \qquad
    {\p}_{\overline{z}} = \tfrac{1}{2} \bigl( \p_\sigma + i \, \p_\tau \bigr)\,.
\ee
The worldsheet fields are the worldsheet scalars parametrizing the spacetime coordinates $X^\mu = (X^A\,, X^{A'})$\,, with $A = 0, 1$ and $A' = 2\,, \cdots\,, d-1$\,,  where $d$ denotes the spacetime dimension, and two additional one-form fields $\lambda$ and $\overline{\lambda}$. The fields $X^A$ 
parametrize longitudinal coordinates and $X^{A'}$ transverse coordinates. The Riemann surface $\Sigma$ is parametrized by the Euclidean coordinates $\sigma^\alpha = (\tau, \sigma)$.   The critical dimension is $d=26$ and $d=10$ for   bosonic and supersymmetric nonrelativistic string theory.

Let us consider open string theory by analyzing the  boundary conditions for the worldsheet fields on a surface $\Sigma$ with a boundary $\partial \Sigma$ at  $\sigma=0$\,. 
Taking variations with respect to the worldsheet fields in \eqref{eq:Snonrel} and requiring the bulk equations of motion,   
\begin{subequations}
\begin{align}
{\partial}_{\overline{z}} X=\partial_z \overline{X}&=0\,,\\
\partial_{\overline{z}} \lambda =\partial_z \overline{\lambda}&=0\,,\\
  \partial_z \partial_{\overline{z}} X^{A'} & = 0\,,
\end{align}
\end{subequations}
to hold, we find the following boundary contribution:
\begin{align}
    \delta S_\text{flat} & = 
    \frac{1}{2\pi\alpha'} \int_{\p \Sigma} d \tau \ls \delta X^{A'} \p_\sigma X^{A'} + \tfrac{1}{2} (\lambda + \overline{\lambda} ) \, \delta X^0 +\tfrac{1}{2}(\lambda-\overline{\lambda})\,\delta X^1 \rs.
    \label{boundaryva}
\end{align}
We impose a Dirichlet boundary condition in the  longitudinal spatial $X^1$-direction \cite{Danielsson:2000mu}\footnote{We can also  consider    boundary conditions   in the presence of a constant longitudinal $B$-field  %
\be
   \frac{1}{4\pi\alpha'} \int_\Sigma d^2 \sigma   \, \epsilon^{\alpha\beta} \, \p_\alpha X^A \, \p_\beta X^B \, \epsilon_{AB} = - \frac{1}{2\pi\alpha'} \int_{\p\Sigma} d \tau \, X^0 \, \p_\tau X^1\,.
\ee
This term is a total derivative and it vanishes when the Dirichlet boundary condition in \eqref{eq:flatBC1} is introduced, but otherwise leads to \cite{Gomis:2000bd} noncommutative open string theory \cite{Seiberg:2000ms,Gopakumar:2000na} on a spacetime-filling brane.}  
\be \label{eq:flatBC1}
     X^1 \Big|_{\sigma = 0} = X^1_0\,,
\ee
where $X^1_0$ is constant. This implies $\delta X^1 |_{\sigma = 0} = 0$\,.
Furthermore, varying with respect to $X^0$ gives rise to the   following boundary condition for the  one-form fields
\be \label{eq:flatBC3}
    \lambda + \overline{\lambda}\,\Big|_{\sigma = 0} = 0\,.
\ee
The equations of motion of $X$ and $\overline{X}$ imply that $X^0$ obeys a Neumann boundary condition   
\be \label{eq:flatBC4}
    \p_\sigma X^0 \Big|_{\sigma = 0} = 0\,,
\ee
and that\footnote{Equation  (\ref{eq:flatBC5}) relates the derivatives $\p_\sigma X^1$ and $\p_\tau X^0$ at the boundary of the worldsheet, and will
play an important role in \S\ref{sec:vertex} when studying the space of open string vertex operators.}
\be \label{eq:flatBC5}
    \p_\sigma X^1 + i \, \p_\tau X^0 \Big|_{\sigma = 0} = 0\,.
\ee
 The transverse coordinates behave the same  as in 
conventional   string theory, and henceforth we consider Neumann boundary conditions for   the transverse coordinates  $X^{A'}$
\be\label{eq:flatBC2}
     \p_\sigma X^{A'} \Big|_{\sigma = 0} = 0\,.
\ee
These boundary conditions preserve  conformal invariance since \eqref{eq:flatBC1} and \eqref{eq:flatBC4} implies that $(\partial_z X+\p_{\overline{z}} \overline{X} )\big|_{\sigma = 0} = 0$\,, which  with  \eqref{eq:flatBC3} and \eqref{eq:flatBC2}  says that $(T -\overline{T})\big|_{\sigma = 0}=0$.\footnote{See \eqref{stressenergy} for the formula for the worldsheet stress energy   tensor $T$.}
These boundary conditions define  the open string theory on a D$(d$$-$2)-brane that is transverse to the longitudinal spatial $X^1$-direction.

The worldsheet theory    (\ref{eq:Snonrel}) with    these boundary conditions defines nonrelativistic open string theory in flat spacetime, which has an open string spectrum with a Galilean invariant dispersion relation. The nonrelativistic open string energy spectrum is   \cite{Danielsson:2000mu} 
\be \label{eq:disre}
    p^{}_0 = \alpha'\, \frac{ p^{}_{A'} p^{}_{A'}}{2 w R} + \frac{N_{\text{open}}}{2 w R}\,,
\ee
where the $X^1$-direction is taken to be  a circle of radius $R$\,.\footnote{We can also consider   open strings   stretched between D-branes separated by a distance $L$  in the $X^1$ direction. Then 
$w R$ in \eqref{eq:disre} must be replaced with $L / (2\pi)$\,.}  We denote  the winding number along $X^1$ as $w$\,, the open string excitation number as $N_{\text{open}}$\,, and the transverse spacetime momentum as $p_{A'}$\,.
 Since the spectrum \eqref{eq:disre} is singular when $w=0$\,, all asymptotic states must necessarily have nonzero winding (unless $p_{A'}=0$). However, off-shell states in the zero winding sector   play the role of intermediate states that mediate an instantaneous electromagnetic force between winding strings, akin to   closed string states with zero winding mediating instantaneous gravitational forces. 
 
One can nevertheless deform the worldsheet theory  (\ref{eq:Snonrel}) with open string vertex operators with $w=0$.   This changes the background fields in which nonrelativitsic open strings propagate. In this paper, we determine the open string backgrounds on which nonrelativistic string theory can be consistently defined.  In the following, we   mostly focus on the zero winding sector. We will discuss nonzero windings   in \S\ref{sec:winding}.

\subsection{Bargmann Symmetry on D-brane} \label{sec:symmetries}
 
Now, we discuss the global symmetries of  the nonrelativistic action in flat space \eqref{eq:Snonrel}. When the surface $\Sigma$ has no boundary,
  the action  \eqref{eq:Snonrel} is invariant under the following infinitesimal global symmetry transformations:
\begin{subequations} \label{eq:dwitt}
\begin{align}
    \delta X^{A'} & = g^{A'} (X) + \overline{g}^{A'} (\overline{X}) - \Lambda^{A'}{}_{B'} X^{B'}, \\[2pt]
    \delta X & = f (X)\,,
        \qquad
    \delta \lambda = - \lambda \, \p^{}_X f(X) - 2 \, \p^{}_X g_{A'} (X) \, \p X^{A'}, \\[3pt]
    \delta \overline{X} & = \overline{f} (\overline{X})\,,
        \qquad
    \delta \overline{\lambda} = - \overline{\lambda} \, \p^{}_{\overline{X}} \overline{f} (\overline{X}) - 2 \, \p_{\overline{X}} \, \overline{g}_{A'} (\overline{X}) \, \overline{\p} X^{A'}.
\end{align}
\end{subequations}
These transformations form the so-called ``extended Galilean symmetry algebra," which contains two copies of the Witt algebra \cite{Batlle:2016iel}. This symmetry algebra  reduces  to the string Newton-Cartan algebra when coupling  nonrelativistic string theory to general   background fields (see \S\ref{sec:closed}).

Let us now determine the  symmetries preserved by a  D$(d-2)$-brane. This requires finding the transformations  \eqref{eq:dwitt} that preserve  the boundary conditions 
\eqref{eq:flatBC1}$-$\eqref{eq:flatBC2}  on the  boundary   of the worldsheet. 
First, these conditions impose that
\be
    \p^{}_X g^{A'} (X) = \p^{}_{\overline{X}} \, \overline{g}^{A'} (\overline{X})\,, 
        \qquad
    \p^{}_X f (X) = \p^{}_{\overline{X}} \overline{f} (\overline{X})\, 
\ee
for arbitrary $X$ and $\overline{X}$. This implies that $g^{A'}, \overline{g}^{A'}, f$ and $\overline{f}$ are functions at most linear in their arguments. Moreover, imposing that the transformation
preserves the Dirichlet boundary condition along $X^1$ correlates the constant part of $f$ with that of $\overline{f}$. Thus,  the infinitesimal  symmetry transformations 
 of nonrelativistic open string theory are:
\begin{subequations} \label{eq:Bargmanndinf}
\begin{align}
    \delta X^{A'} & = \Xi^{A'} + \Lambda^{A'} \, X^0 - \Lambda^{A'}{}_{B'} X^{B'}, \\[2pt]
    \delta X^0 & =  \Xi^0 + \Theta \, X^0, 
        \qquad
    \delta \lambda = - \Theta \, \lambda - 2 \Lambda_{A'} \p_z X^{A'}, \\[2pt]
    \delta X^1 & = \Theta \, X^1,
        \qquad\qquad
    \delta \overline{\lambda} = - \Theta \, \overline{\lambda} - 2 \Lambda_{A'} \p_{\overline{z}} X^{A'}.
\end{align}
\end{subequations}
Evaluated on the worldsheet boundary, we have 
\be
    \delta \lambda = - \delta \overline{\lambda} = - \Theta \, \lambda + i \Lambda_{A'} \p_\tau X^{A'}\,.
\ee 
The conserved charges associated to the   different  transformations are:
\begin{subequations}
\begin{align}
    \text{time translation} \quad \Xi^0 \! : & \quad\,\, H = \int d\sigma \, \pi_0\,, \\[2pt]
    \text{transverse translation} \quad \Xi^{A'} \! : & \quad P_{A'} = \int d\sigma \, \pi_{A'}\,, \\[2pt]
    \text{Galilean boost} \quad \Lambda^{A'} \! : & \quad G_{A'} = \int d\sigma \Bigl( X^0 \, \pi_{A'} - \frac{i}{2\pi\alpha'} \, X^{A'} \, \p_\sigma X^1 \Bigr)\,, \\[2pt]
    \text{transverse rotation} \quad \Lambda^{A'}{}_{B'} \! : & \quad J_{A'B'} = \int d\sigma \, \Bigl( - X^{A'} \pi_{B'} + X^{B'} \pi_{A'} \Bigr)\,, \\[2pt]
    \text{longitudinal dilatation} \quad \Theta : & \quad D = \int d\sigma \, X^A \, \pi_{A}\,, \label{eq:dilatation}
\end{align}
\end{subequations}
where we defined the conjugate momentum for $X^0$\,, $X^1$ and $X^{A'}$ as
\be
    \pi_0 = \frac{i}{4\pi\alpha'} \lr \lambda - \overline{\lambda} \rr,
        \qquad
    \pi_1 = \frac{i}{4\pi\alpha'} \lr \lambda + \overline{\lambda} \rr,
        \qquad
    \pi_{A'} = \frac{1}{2\pi\alpha'} \, \p_\tau X^{A'}.
\ee
Using the equal-$\tau$ Poisson brackets,
\begin{align}
    [ X^A (\sigma_1)\,, \pi_B (\sigma_2) ] = \delta^A_B \, \delta (\sigma_1 - \sigma_2)\,,
        \qquad
    [ X^{A'} (\sigma_1)\,, \pi_{B'} (\sigma_2) ] = \delta^{A'}_{B'} \, \delta (\sigma_1 - \sigma_2)\,,
\end{align}
we find that
the generators $H$\,, $P_{A'}$, $G_{A'}$ and $J_{A'B'}$ satisfy the Bargmann algebra,
\begin{subequations} \label{eq:Bargmann}
\begin{align}
    [H, G_{A'}] & = - P_{A'}\,,
        \qquad\,\,\,\,\,\,%
    [P_{A'}, J_{B'C'}] = \delta_{A'B'} \, P_{C'} - \delta_{A'C'} \, P_{B'}\,, \\[2pt]
    [P_{A'}, G_{B'}] & = \delta_{A'B'} \, Z\,,
        \qquad\,%
    [G_{A'}, J_{B'C'}] = \delta_{A'B'} \, G_{C'} - \delta_{A'C'} \, G_{B'}\,, \\[2pt]
    [J_{A'B'}, J_{C'D'}] & = \delta_{B'C'} J_{A'D'} - \delta_{A'C'} J_{B'D'} + \delta_{A'D'} J_{B'C'} - \delta_{B'D'} J_{A'C'}\,,
\end{align}
\end{subequations}
where
\be
    Z = \frac{i}{2\pi\alpha'} \int d\sigma \, \p_\sigma X^1
\ee
is the central charge in the Bargmann algebra. Note that $Z$ measures the winding number of strings along $X^1$. The appearance of the Bargmann algebra here is expected, as the open string theory defines a QFT living on the D-brane, which describes dynamics of nonrelativistic gauge fields that we will elaborate on in the rest of the section.

This symmetry algebra is further extended by  the longitudinal dilatation generator in \eqref{eq:dilatation}, which satisfies the following Lie brackets:
\be \label{eq:do}
    [D\,, H] = H\,,
        \qquad
    [D\,, G_{A'}] = - G_{A'}\,,
        \qquad
    [D\,, Z] = - Z\,.
\ee
Note that the compactified circle in $X^1$ is rescaled under the dilatation transformation. This rescaling can be compensated when a dilaton is included. Consequently, as we will see later at the end of \S\ref{sec:GED}, the dilatation is only a symmetry of the spacetime equations of motion instead of the spacetime action, unless a dilaton background field is present. Also note that the dilatation generator  in \eqref{eq:dilatation} gives rise to the dilaton background in string theory when the algebra is gauged.\footnote{See \cite{Bergshoeff:2019pij} for relevant discussion of the analogue of this dilatation generator in string Newton-Cartan algebra.}

\subsection{Open String Vertex Operators}\label{sec:vertex}

Our next goal is to consider the most general deformation of nonrelativistic open string theory in flat spacetime by perturbing  around the free action \eqref{eq:Snonrel} with open string vertex operators.  We consider here vertex operators with zero winding along $X^1$. Turning on these vertex operators changes the background fields in which nonrelativistic open strings propagate.

In order to classify the open string vertex operators that can be added to the sigma model action \eqref{eq:Snonrel}, we first consider all possible $(1,0)$ and $(0,1)$ forms in the bulk,
\be \label{eq:ofos}
    \lambda\,,
        \quad
    \p_z X\,,
        \quad
    \p_z X^{A'}\,;
        \quad
    \overline{\lambda}\,,
        \quad
    \p_{\overline{z}} \overline{X}\,,
        \quad
    \p_{\overline{z}} X^{A'}.
\ee
In \S\ref{sec:nonrelost}, we derived the boundary conditions in nonrelativistic open string theory with a flat target space (and with zero open string background fields), which we collect as follows:
\begin{subequations}
\begin{align}
    \text{Dirichlet:\quad} & \p_\tau X^1 \Big|_{\sigma = 0} = 0 \,, \\[2pt]
    \text{Neumann:\quad} & \p_\sigma X^0 \Big|_{\sigma=0} = \p_\sigma X^{A'} \Big|_{\sigma=0} = 0\,,
\end{align}
\end{subequations}
and
\begin{align}
    \lambda + \overline{\lambda} \, \Big|_{\sigma = 0} = \p_\sigma X^1 + i \,  \p_\tau X^0 \Big|_{\sigma=0} = 0\,.
\end{align}
Therefore, the   independent one-forms in \eqref{eq:ofos} are
\be
    \p_\tau X^0,
        \quad
    \p_\tau X^{A'},
        \quad
    \lambda\,.
\ee
The most general deformation created by    an open string vertex operator with zero winding is    
\be \label{eq:Vbdry}
    \CV = \int_{\p\Sigma} d\tau \Bigl[ : \! N \lambda : + \, i \bigl( : \! A_0 \, \p_\tau X^0 \! : + : \! A_{A'} \, \p_\tau X^{A'} \! \! : \bigr) \Bigr]. 
\ee
The worldsheet couplings $N$, $A_0$ and $A_{A'}$ are functions of $X^0$ and $X^{A'}$ and have a spacetime interpretation as background open string fields on the D$(d-2)$-brane worldvolume. 
Here, $A_0$ and $A_{A'}$ are the components of a $U(1)$ gauge field.  The vertex operator \eqref{eq:Vbdry} is invariant under the $U(1)$ gauge transformation
\be \label{eq:gaugetrnsf}
    \delta_\epsilon A_0 = \p_0 \epsilon\,,
        \qquad %
    \delta_\epsilon A_{A'} = \p_{A'} \epsilon\,.
\ee
The coupling $N = N (X^0, X^{A'})$, which is invariant under the $U(1)$ gauge symmetry, has a geometrical interpretation as the Nambu-Goldstone boson that perturbs around the solitonic D$(d-2)$-brane, which breaks the translational symmetry along the $X^1$ direction.

The   vertex operator \eqref{eq:Vbdry} is invariant under the global spacetime symmetry algebra we derived in the previous section when supplementing the 
worldsheet transformations in \eqref{eq:Bargmanndinf}
 with the following  infinitesimal  transformations of the background fields\,:
\begin{subequations} \label{eq:NA0AA'}
\begin{align}  
    \delta N  & = - \xi^0 \, \p_0 N - \xi^{A'} \p_{A'} N + \Theta N\,, \\[2pt] 
    \delta A_{0} & = - \xi^0 \, \p_0 A_0 - \xi^{A'} \p_{A'} A_0 - A_0 \, \p_0 \xi^0 - A_{A'} \p_0 \xi^{A'}, \\[2pt]
    \delta A_{A'} & = - \xi^0 \, \p_0 A_{A'} - \xi^{B'} \p_{B'} A_{A'} - A_0 \, \p_{A'} \xi^0 - A_{B'} \p_{A'} \xi^{B'} - \Lambda_{A'} N\,,
\end{align}
\end{subequations}
where 
\be
    \xi^0 \equiv \Xi^0 + \Theta X^0\,, 
        \qquad
    \xi^{A'} \equiv \Xi^{A'} + \Lambda^{A'} X^0 - \Lambda^{A'}{}_{B'} X^{B'}.
\ee
In particular, the transformations under the Galilean  boost symmetry are
\begin{subequations}
\begin{align}
    \delta_G N &= - \Lambda^{A'} X^0 \, \p_{A'} N\,, \\[2pt]
    \delta_G A_0 & = - \Lambda^{A'} X^0 \, \p_{A'} A_0 - \Lambda^{A'} A_{A'}\,, \\[2pt]
    \delta_G A_{A'} & = - \Lambda^{B'} X^0 \, \p_{B'} A_{A'} - \Lambda_{A'} N\,. 
\end{align}
\end{subequations}
These transformations realize on the background fields $(A_0,A_{A'},N)$ the Bargmann algebra (with the additional dilatation symmetry) that is preserved by the D-brane. 
The expression in \eqref{eq:Vbdry} is the most general open string vertex operator with zero winding respecting these symmetries. 

Our goal is to determine the open string backgrounds in which nonrelativistic open strings in flat spacetime can propagate on.

\subsection{Galilean Electrodynamics from Nonrelativistic Open String Theory} \label{sec:GED}

Requiring that the open string vertex operator in \eqref{eq:Vbdry} is BRST invariant imposes constraints on the couplings $(A_0,A_{A'},N)$. These conditions are the linearized spacetime equations of motion that the background fields $(A_0,A_{A'},N)$ must satisfy for a self-consistent nonrelativistic open string theory. 

The (holomorphic) BRST charge of nonrelativistic string theory \eqref{eq:Snonrel} is
\be
    Q = \int \frac{dz}{2\pi i} \ls : \! c \, T^m (z) \! : + : \! b \, c \, \p_z c \! : + \, \tfrac{3}{2} : \! \p_z^2 c \! : \rs\!, 
\ee
where the matter stress energy tensor is
\begin{align}
\label{stressenergy}
    T^m (z) & = - \frac{1}{2\alpha'} \lr 2 \, \p_z X^{A'} \, \p_z X^{A'} + \lambda \, \p_z X \rr\!.
\end{align}
Our goal is to calculate the BRST transformation of the most general open string vertex operator \eqref{eq:Vbdry}.

Calculating the BRST  transformation requires computing operator product expansions (OPEs). Using the Neumann boundary condition \eqref{eq:flatBC2} for $X^{A'}$\,, we find
\be \label{eq:XX}
    X^{A'} (z, \overline{z}) \, X^{B'} (z', \overline{z}') \sim - \frac{1}{2} \, \alpha' \, \delta^{A'B'} \bigl( \, \ln |z-z'|^2 + \ln |z + \overline{z}'|^2 \, \bigr)\,.
\ee
Moreover, the OPEs for the holomorphic fields $\lambda = \lambda (z)$ and $X = X(z)$\,, and the anti-holomorphic fields $\overline{\lambda} (\overline{z})$ and $\overline{X} (\overline{z})$ are
\be \label{eq:lambdaX}
    \lambda (z) \, X (z') \sim - \frac{2\alpha'}{z - z'}\,,
        \qquad%
    \overline{\lambda} (\overline{z}) \, \overline{X} (\overline{z}') \sim - \frac{2\alpha'}{\overline{z} - \overline{z}'}\,.
\ee
We decompose the worldsheet field $X^\mu$ into its left- and right-moving parts as
\be
    X^\mu (z\,, \overline{z}) = X_L^\mu (z) + X_R^\mu (\overline{z})\,.
\ee
In doing so, the (anti-)holomorphic equations of motion $\overline{\p} X = \p \overline{X} = 0$ implies
\be
    X (z) = 2 \, X_L^0 (z) = 2 \, X_L^1 (z)\,,
        \qquad%
    \overline{X} (\overline{z}) = 2 \, X_R^0 (\overline{z}) = - 2 \, X_R^1 (\overline{z})\,.
\ee
In what follows, we focus on the left-moving modes $X_L^{A'} (z)$\,, $X_L^0 (z)$, and $\lambda(z)$\,, with
\begin{align}
    X_L^{A'} (z) \, X_L^{B'} (z') \sim - \frac{1}{2} \, \alpha' \, \delta^{A'B'} \ln (z- z')\,,
        \qquad
    \lambda (z) \, X_L^0 (z') \sim - \frac{\alpha'}{z - z'}\,.
\end{align}

In terms of the holomorphic fields and the holomorphic coordinate $z$\,, we write the open string vertex operator $\CV$ in \eqref{eq:Vbdry} as
\be
    \CV = \int_{\p\Sigma} dz \, V_L (z) - \int_{\p\Sigma} d\overline{z} \, V_R (\overline{z})\,.
\ee
 For open strings, the holomorphic and anti-holomorphic parts are related to each other by the boundary conditions, so it is sufficient to focus on either one of them. In what follows, we will study the holomorphic part, 
\be \label{eq:mathcalVL}
    \CV_L = \int_{\p\Sigma} dz \, V_L(z)\,,
\ee
 with 
\be \label{eq:VLz}
    V_L (z) =
    \tfrac{1}{2} : \! N \, \lambda (z) \! : - : \! A_0 \, \p_z X^0_L (z) \! : - : \! A_{A'} \, \p_z X_L^{A'} \!(z) \! : .
\ee
It then follows that the BRST transformation of $V_L$  is
\be
    [Q\,, V_L (z)] = \, : \! \p_z W_L (z) \! : - \frac{\alpha'}{4} \! : \! \p_z c (z) \, \mathcal{E} (z) \! :\,,
\ee
where
\begin{subequations}
\begin{align}
    W_L (z) & = c(z) \, V_L (z) + \frac{\alpha'}{4} \, \p_z c \lr \p^{A'} \! A_{A'} - \p_0 N \rr, \\[2pt]
    \mathcal{E} (z) & = \frac{1}{2} \, \p^{A'} \p_{A'} N \, \lambda - \lr \p_0^2 N - \p^{A'} \! E_{A'} \rr \p X^0 - \lr \p_{0} \p_{A'} N + \p^{B'} \! F_{B'A'} \rr \p X^{A'}, \label{eq:CE}
\end{align}
\end{subequations}
and we have introduced the gauge invariant electric and magnetic fields,
\be
    E_{A'} = \p_0 A_{A'} - \p_{A'} A_0\,,
        \qquad
    F_{A'B'} = \p_{A'} A_{B'} - \p_{B'} A_{A'}\,.
\ee
Requiring that $\CV_L$ in \eqref{eq:mathcalVL} is BRST invariant implies that $[Q, V_L(z)]$ must be  a total derivative, which   requires that
$\mathcal{E} = 0$\,.
This sets the couplings in front of the operators $\lambda$\,, $\p X^0$ and $\p X^{A'}$ in \eqref{eq:CE} to zero separately, %
\begin{subequations} \label{eq:eomGED}
\begin{align} 
    \p^{A'} \! \p_{A'} N & = 0\,, \\[2pt]
    \p_0^2 N - \p^{A'} \! E_{A'} & = 0\,, \\[2pt]
        \qquad
    \p_{A'} \p_0 N + \p^{B'} \! F_{B'A'} & = 0\,.
\end{align}
\end{subequations}
Therefore,  when the background fields   $(A_0,A_{A'},N)$ satisfy the linearized equations of motion \eqref{eq:eomGED},   nonrelativistic 
open string theory can be consistently defined. 

The linearized background field equations in \eqref{eq:eomGED} arise from varying the following action with respect to $N$\,, $A_{0}$ and $A_{A'}$\,,
\be \label{eq:GEDflat}
    S_\text{} = \frac{1}{g^2} \int dX^0 \, dX^{A'} \Bigl( \tfrac{1}{2} \, \p_0 N\,\p_0 N - E_{A'}\p_{A'}N - \tfrac{1}{4}F_{A'B'}F_{A'B'} \Bigr)\,,
\ee 
where $g^2=e^{\Phi_0}$, with $\Phi_0$ the dilaton expectation value. 
This theory is known   as Galilean Electrodynamics in the literature, which was historically discovered by considering a null reduction of Maxwell theory in \cite{Santos:2004pq}, and later reproduced as a nonrelativistic limit of Maxwell theory with a free massless scalar in \cite{Bergshoeff:2015sic, Festuccia:2016caf}.\footnote{In \cite{Festuccia:2016caf}, a finite Galilean boost transformation was introduced, under which \eqref{eq:GEDflat} is invariant. In the worldsheet formalism we are working with, if one takes the boost transformation $X^A \rightarrow X^A$ and $X^{A'} \rightarrow X^{A'} + \Lambda^{A'} X^0$ in \eqref{eq:Bargmanndinf} to be a finite transformation, then, by requiring the nonrelativistic string  action \eqref{eq:Snonrel} to be invariant, the boost transformation of the one-form fields $\lambda$ and $\overline{\lambda}$ gain a term quadratic in $\Lambda_{A'}$\,, 
\be
    \lambda \rightarrow \lambda - \Lambda_{A'} \p X^{A'} - \tfrac{1}{2} \, \Lambda_{A'} \Lambda^{A'} \p X^0\,,
        \qquad
    \overline{\lambda} \rightarrow \overline{\lambda} - \Lambda_{A'} \overline{\p} X^{A'} - \tfrac{1}{2} \, \Lambda_{A'} \Lambda^{A'} \overline{\p} X^0\,.
\ee
Requiring the open string vertex operator \eqref{eq:Vbdry} to be invariant under these transformations gives
\begin{subequations}
\begin{align}
    N'(X') = N(X)\,, 
        \qquad%
    A'_0 (X') & = A_0 (X) - \Lambda^{A'} \! A_{A'}(X) + \tfrac{1}{2} \, \Lambda_{A'} \Lambda^{A'} N(X)\,, \\[2pt]
    A'_{A'} (X') & = A_{A'} (X) - \Lambda^{A'} N,
\end{align}
\end{subequations}
which derives the finite boost transformations given in \cite{Festuccia:2016caf}.}
The action \eqref{eq:GEDflat} is invariant under  the Bargmann symmetry, and if we appropriately shift the dilaton,  also under the dilatation symmetry \eqref{eq:dilatation}, that is, dilatation transformations are a symmetry of the equations of motion but not of the action \eqref{eq:GEDflat}. 

As mentioned earlier, in the zero winding sector, there are no propagating degrees of freedom. Nevertheless, strings with no winding can appear as intermediate states that mediate instantaneous electromagnetic forces between winding strings. The fact that open strings with no winding do not have propagating degrees of freedom is realized in the absence of propagating modes in the Galilean Electrodynamics action \eqref{eq:GEDflat}.

In \S\ref{sec:curved}, we   generalize the Galilean Electrodynamics result in \eqref{eq:eomGED} to its DBI analogue in    arbitrary open and closed background fields.  

\subsection{Galilean Yang-Mills Theory from Nonrelativistic Open String Theory}\label{sec:YM}

Our next goal is to derive the low energy effective action that determines the  consistent open string backgrounds on a stack of $n$ coincident  D$(d-2)$-branes. This requires introducing $U(n)$ Chan-Paton bundles. The most general open string vertex operator with vanishing winding number is described by the Wilson line like insertion
\be \label{eq:W}
    W = \tr \, \mathcal{P} \exp \lc \int_{\p\Sigma} d\tau \ls : \! N \lambda \! : + i \bigl( : \! A_0 \, \p_\tau X^0 \! : + : \! A_{A'} \, \p_\tau X^{A'} \!\! : \bigr) \rs \rc,
\ee
where the path-ordering operator $\CP$ orders terms such that higher values of $\tau$ stand to the left.  We defined the adjoint fields 
\be \label{eq:A0A'NU(N)}
    A_0 = A_0^m \, T^m,
        \qquad
    A_{A'} = A_{A'}^m \, T^m,
        \qquad
    N = N^m \, T^m,
\ee
where $T^m$ are generators in the fundamental representation of $U(n)$\,.  There are  $n^2$ generators $T^m$\,, each of which is a hermitian $n \times n$ matrix.  The fields in \eqref{eq:A0A'NU(N)} transform under $U(n)$ gauge transformations as follows:
\begin{subequations}
\begin{align} 
    A_{0} & \rightarrow U A_{0} \, U^\dagger + i \, U (\p_0 U^\dagger)\,, \\[2pt]
    A_{A'} & \rightarrow U A_{A'} \, U^\dagger + i \, U (\p_{A'} U^\dagger)\,, \\[2pt]
      N &\rightarrow U N \, U^\dagger\,, 
\end{align}
\end{subequations} 
where 
\be
    U (X_0, X_{A'}) = \exp\ls i \, \alpha^m (X_0, X_{A'}) \, T^m\rs\,.
\ee
These leave the vertex operator \eqref{eq:W} invariant. 
$N$ is therefore a field in the adjoint representation of $U(n)$. 
The gauge-covariant field strengths are
\begin{subequations}
\begin{align}
    E_{A'} & = \p_0 A_{A'} - \p_{A'} A_0 - i \, [A_0\,, A_{A'}]\,, \\[2pt]
    F_{A'B'} & = \p_{A'} A_{B'} - \p_{B'} A_{A'} - i \, [A_{A'}, A_{B'}]\,.
\end{align}
\end{subequations}
We define the covariant derivatives $D_0$ and $D_{A'}$\,, which act on the adjoint scalar $N$ as
\be
    D_{0} N = \p_0 N  - i \, [A_0, N]\,, 
        \qquad
    D_{A'} N = \p_{A'} N  - i \, [A_{A'}, N]\,.
\ee
We would like to derive the Yang-Mills analogue of the Galilean Electrodynamics equations of motion in \eqref{eq:eomGED}, which involves keeping track of terms up to the third order in  the background fields in \eqref{eq:W} in the following expansion:
\begin{align} 
    W = \tr \biggl[ \mathbb{1} & + \int_{-\infty}^\infty d\tau \, V (\tau) + \int_{-\infty}^\infty d\tau \, V (\tau) \int_{-\infty}^\tau d\tau' \, V (\tau') \notag \\[2pt]
    & + \int_{-\infty}^\infty d\tau \, V (\tau) \int_{-\infty}^\tau d\tau' \, V (\tau') \int_{-\infty}^{\tau'} d\tau''\, V(\tau'') + \cdots \biggr],
\end{align}
where
\be
    V = :\!N \lambda:\! + \, i \bigl( :\!A_0 \, \p_\tau X^0\!: + :\!A_{A'} \, \p_\tau X^{A'}\!\!: \bigr).
\ee
In principle, the gauge covariant equations of motion can be derived by requiring the BRST invariance of $W$ order by order, which in practice is quite complicated. We will therefore use a slightly different method to extract higher-order contributions of the gauge fields to the equations of motion, following closely \cite{Polchinski:1998rq, Gomis:2019zyu}.   

We start with introducing an Einbein field $e(s)$ with $d\tau = ds \, e(s)$ in the measure of the boundary action.  This field transforms under   Weyl symmetry as $\delta_W e = \delta \omega \, e$\,. 
Moreover, the one-form field also transforms nontrivially under the Weyl symmetry, with $\delta_W \lambda = - \delta \omega \, \lambda$\,.
Then, the scalar-coupled Wilson line can be written as
\be
    W = \tr \, \mathcal{P} \exp \lc \int_{\p\Sigma} ds \, e \ls N \lambda + \frac{i}{e} \lr A_0 \, \p_s X^0 + A_{A'} \, \p_s X^{A'} \rr \rs \rc,
\ee
which is Weyl invariant classically. However, the renormalization of a classically Weyl invariant operator may lead to a Weyl anomaly. To see this, we define a renormalized operator $[\mathcal{O} (\tau)]_r$ for a given boundary operator $\CO (\tau)$\,, 
\be
    [\CO]_r = \exp \! \ls \alpha' \! \! \int \! d s_1 \, ds_2 \, \ln d(s_1, s_2) \, \widehat{P} (s_1\,, s_2) \rs \CO\,,
\ee
where
\be
    \widehat{P} (s_1\,, s_2) = \frac{\delta}{\delta X^{A'}(s_1)} \frac{\delta}{\delta X^{A'}(s_2)} - 2 i e^{-1} \p_{s_1} \frac{\delta}{\delta \lambda (s_1)} \frac{\delta}{\delta X^0(s_2)}\,.
\ee
Here, $d(s_1\,, s_2) = \int_{s_1}^{s_2} d\tau$ is the distance between $\tau(s_1)$ and $\tau(s_2)$ on $\p\Sigma$\,, which transforms nontrivially under the Weyl symmetry, satisfying the following coincidence limits \cite{Polchinski:1998rq}:
\be
    \delta_W \ln d(s\,, s') \big|_{s'=s} = \delta \omega(s)\,,
        \qquad
    \delta_W \, \p_s \ln d(s\,, s') \big|_{s'=s} = \tfrac{1}{2} \, \p_\tau \delta \omega(s)\,,
        \quad \cdots
\ee
At the lowest order of the field strength and $N$, we find
\be
    \delta_W [W]_r = \alpha' \!\! \int_{\p\Sigma} ds \, e \Bigl[ \lr \lambda \, \delta_W N + i e^{-1} \bigl( \, \delta_W A_0 \, \p_s X^0 + \delta_W A_{A'} \, \p_s X^{A'} \bigr) \! + \cdots \rr W \Big]{}^{}_r\,,
\ee
where
\begin{subequations}
\begin{align}
    \delta_W N &= \delta \omega \, D_{A'} D^{A'} \! N, \\[2pt]
    \delta_W A_0 & = \delta \omega \lr D_0^2 N - D^{A'} E_{A'} \rr, \\[2pt]
    \delta_W A_{A'} & = \delta \omega \lr D_{0} D_{A'} N + D^{B'} \! F_{B'A'} \rr.
\end{align}
\end{subequations}
Setting $\delta_W [W]_r$ to zero at the lowest order in the field strength gives rise to the equations of motion\footnote{See \cite{Bagchi:2015qcw} for other versions of Yang-Mills theories with distinct nonrelativistic symmetry.} of a  $U(n)$ Yang-Mills theory with nonrelativistic symmetry
\begin{subequations}
\begin{align}
    D_{A'} D^{A'} N&=0\,, \\[2pt]
  D_0^2 N - D^{A'} E_{A'} & =0 \,, \\[2pt]
  D_{0} D_{A'} N + D^{B'} \! F_{B'A'}&=0\, .
\end{align}
\end{subequations}
These equations of motion can be derived from the following nonrelativistic Yang-Mills action:
 \be 
    S_\text{YM} = \frac{1}{g^2_\text{YM}} \int dX^0 \, dX^{A'} \, \tr \Bigl( \tfrac{1}{2} \, D_0 N\,D_0 N - E_{A'}D_{\!A'}N - \tfrac{1}{4}F_{A'B'}F_{A'B'} \Bigr)\,.
\ee 
This nonlinear, nonabelian gauge theory has   Bargmann symmetry. The corresponding equations of motion, but not the action, are invariant under the dilatation transformation \eqref{eq:dilatation}.

\subsection{Winding Open String Backgrounds}
\label{sec:winding}

We  now consider turning on open string vertex operators with nonzero winding number. This introduces novel and interesting elements in the discussion, like an emergent 
coordinate conjugate to the winding number, that leads to a mild nonlocality in the spacetime equations of motion, which now describe propagating degrees of freedom.

We  first analyze vertex operators with a fixed winding number $w$. Focusing on the holomorphic part, the open string vertex operator \eqref{eq:VLz} is 
\be
    V^w_L (z) = \, : \! \ls \tfrac{1}{2} N^w \lambda (z) - A^w_0 \, \p_z X^0_L (z) - A^w_{A'} \p_z X^{A'}_L (z) \rs e^{i q \! \int^z \! dz' \lambda(z')} \! :\,,
\ee
where $q$ encodes the winding number $w$,
\be
    q =  \frac{w R}{2 \alpha'}\,,
\ee
with $R$ the radius of the compactified $X^1$-direction.  The new worldsheet couplings $N^w$, $A^w_0$, and $A^w_{A'}$ are related to $N$, $A_0$\,, and $A_{A'}$ by a ``Fourier transform"
\begin{subequations}
\begin{align}
    N (X^0, X^{A'}, \lambda) & = \sum_{w} N^w (X^0, X^{A'}) \, e^{i q \! \int^z \! dz' \lambda(z')}, \\[2pt]
    A_0 (X^0, X^{A'}, \lambda) & = \sum_{w} A_0^w (X^0, X^{A'}) \, e^{i q \! \int^z \! dz' \lambda(z')}, \\[2pt]
    A_{A'} (X^0, X^{A'}, \lambda) & = \sum_w A_{A'}^w (X^0, X^{A'}) \, e^{i q \! \int^z \! dz' \lambda(z')}.
\end{align}
\end{subequations} 
The integrated vertex operator $\mathcal{V}^w_L = - 2 \int_{\p\Sigma} dz \, V^w_L(z)$ is invariant under the $U(1)$ gauge transformations 
\be
    \delta_\epsilon N^w = - 2 i q \, \epsilon\,,
        \qquad
    \delta_\epsilon A^w_0 = \p_0 \epsilon\,,
        \qquad
    \delta_\epsilon A^w_{A'} = \p_{A'} \epsilon\,.
\ee
We note that winding number endows $N^w$ with a nontrivial $U(1)$ transformation. This fact has a simple interpretation if we  introduce a  novel spacetime  coordinate conjugate to winding (see \eqref{newcoord}). This suggests that the dynamics of wound open strings is more naturally formulated in an extended spacetime, as we shall now see.

The nonrelativistic spacetime transformations also act on $N^w$, $A^w_0$, and $A^w_{A'}$ differently from the way they act on $N$, $A_0$, and $A_{A'}$ in \eqref{eq:NA0AA'}. For example, the boost transformation parametrized by $\Lambda^{A'}$ in \eqref{eq:NA0AA'} is now modified to be
\begin{subequations}
\begin{align}
    \delta_G N^w &= 2i q \, \Lambda_{A'} X^{A'} N - \Lambda^{A'} X^0 \, \p_{A'} N\,, \\[2pt]
    \delta_G A^w_0 & = 2 i q \, \Lambda_{A'} X^{A'} A_0 - \Lambda^{A'} X^0 \, \p_{A'} A_0 - \Lambda^{A'} A_{A'}\,, \\[2pt]
    \delta_G A^w_{A'} & = 2 i q \, \Lambda_{B'} X^{B'} A_{A'} - \Lambda^{B'} X^0 \, \p_{B'} A_{A'} - \Lambda_{A'} N\,. 
\end{align}
\end{subequations}

The BRST transformation of $V_L^w (z)$ is given by
\be \label{eq:QVLw}
    [Q\,, V_L^w (z)] = \, : \! \p_z W_L^w (z) \! : - \frac{\alpha'}{4} : \! \p_z c (z) \, \mathcal{E}^w (z)\,e^{i q \! \int^z \! dz' \lambda(z')}\!:\,,
\ee
with
\begin{subequations}
\begin{align}
    W^w_L (z) & = c(z) \, V^w_L (z) + \frac{\alpha'}{4} \, \p_z c \lr \p^{A'} \! A^w_{A'} - \p_0 N^w  +  2 i q A^w_0 \rr e^{i q \! \int^z \! dz' \lambda(z')}, \label{eq:dc} \\[2pt]
    \mathcal{E}^w (z) & = \frac{1}{2} \lr \p^{A'} \! \p_{A'} N^w + 2 i q \, \p_0 N^w + 2 i q \, \p^{A'} \! A^w_{A'} - 4 q^2 A^w_0 \rr \lambda \notag \\[2pt]
    & \quad - \lr \p_0^2 N^w - \p^{A'} \! E^w_{A'} + 2 i q \, \p_0 A^w_0 \rr \p X^0 \notag \\[2pt]
    & \quad - \lr \p_{0} \p_{A'} N^w + \p^{B'} \! F^w_{B'A'}  + 2 i q \, \p_0 A^w_{A'}  +  2 i q E^w_{A'} \rr \p X^{A'},
\end{align}
\end{subequations}
where we have defined wound electric and magnetic fields 
\be
    E^w_{A'} = \p_0 A^w_{A'} - \p_{A'} A^w_0\,,
        \qquad
    F^w_{A'B'} = \p_{A'} A^w_{B'} - \p_{B'} A^w_{A'}\,.
\ee
 
It is useful  to consider  the vertex operator obtained by summing over all windings,  
\be
 \CV_L = \sum_w \CV_L^w\,.
\ee
Demanding that $\CV_L = \sum_w \CV_L^w$ is BRST invariant sets $\sum_w \mathcal{E}^w = 0$\,, with $\mathcal{E}^w$ given in \eqref{eq:QVLw}. It is enlightening to define an auxiliary coordinate,
\be
    X^r (z) \equiv \frac{1}{2} \int^z dz' \lambda(z')\,, 
    \label{newcoord}
\ee
and introduce the ``spacetime" coordinates     $X^\CA = (X^r, X^0, X^{A'})$. This is reminiscent of a spacetime doubling as in double field theory \cite{Hull:2009mi}.     
Using \eqref{eq:QVLw}, we find that $\sum_w \mathcal{E}^w = 0$ gives rise to a system of equations of motions that admits the following simple representation:
\begin{align} \label{eq:GdF}
    G^{\CB\CC} \p_\CB F_{\CC\CA} = 0,
\end{align}
where $F_{\CA\CB} = \p_{\CA} A_{\CB} - \p_{\CB} A_{\CA}$\,, with $A_\CA = \bigl(-N, A_0, A_{A'} \bigr)$\,, and
\be
    G^{\CA\CB} = 
        \begin{pmatrix}
            0 & \quad 1 & 0 \\[2pt]
            1 & \quad 0 & 0 \\[2pt]
            0 & \quad 0 & \quad \delta^{A'B'}
        \end{pmatrix}.
        \ee
Here, $N$, $A_0$ and $A_{A'}$ are functions of the extended spacetime coordinates $X^\CA$. 

The equations of motion for the open string fields with winding \eqref{eq:GdF} can be derived by varying the following action:
\begin{align} \label{eq:Swinding}
    S_\text{} & = - \frac{1}{4 g^2} \int dX^r\, dX^0 \, dX^{A'} F_{\CA\CB} \, G^{\CA\CC} G^{\CB\CD} F_{\CC\CD}\,. 
\end{align}
We note that we have been able to write down an action by introducing an auxiliary coordinate $X^r$, conjugate to winding number.  
The explicit dependence on the extended direction $X^r$ rather than only the local coordinates $X^0$ and $X^{A'}$ signatures nonlocality. This nonlocal feature is expected as the winding sector is inherently nonlocal. This nonlocality is, however, of a rather tame form, as we are able to write down explicitly the spacetime action \eqref{eq:Swinding} in terms of the auxiliary coordinate $X^r$.
 
In contrast to Galilean Electrodynamics, which is non-dynamical, and which describes the sector with vanishing winding,  the gauge theory in \eqref{eq:Swinding} has $d-2$ propagating degrees of freedom with a nonrelativistic dispersion relation, 
\be
    p_0 = \alpha' \, \frac{p_{A'} p_{A'}}{2 w R}\,.
\ee
This beautifully realizes  the nonrelativistic open string dispersion relation in \eqref{eq:disre} with the open string excitation number $N_{\text{open}}$ set to zero.

The compact form of the equations of motion in \eqref{eq:GdF} finds a simple interpretation if one performs a T-duality transformation along the longitudinal spatial $X^1$-direction, which is compactified on a circle of radius $R$\,. After performing this longitudinal spatial T-duality transformation in nonrelativistic open string theory, we find relativistic open string theory on a spacetime-filling brane background, with the following duality dictionary \cite{Tdualopenstring}: the $X^1$-direction is dual to a lightlike direction $\tilde{X}^1 = X^r$ on the brane, the Goldstone boson $N$ and the gauge component $A_0$ are dual to the lightlike components of the relativistic $U(1)$ gauge field, and the winding number $w$ is dual to the discrete momentum along the lightlike circle.

\vspace{1mm}

This concludes our analysis of nonrelativistic open string theory in flat spacetime. 

\section{Curved Backgrounds} \label{sec:curved}

In this section, we consider the nonrelativistic string
nonlinear sigma model  in an arbitrary string Newton-Cartan geometry, $B$-field, $U(1)$ gauge field, and dilaton background in presence of a  D($d-2$)-brane. Our goal is to compute the beta-functions for the open string couplings, which define the spacetime equations of motion of nonrelativistic open string theory.
 We   focus our discussion  to the zero winding sector. We start with a brief review of the nonrelativistic closed string nonlinear sigma model, and then move on to the construction of the Dirichlet nonlinear sigma model that couples nonrelativistic open strings to a closed string background geometry.

\subsection{Closed Strings on a String Newton-Cartan Background} \label{sec:closed}

The appropriate closed string background geometry for nonrelativistic string theory is string Newton-Cartan geometry, which is a non-Lorentzian and non-Riemannian geometry. We give a short review of the definition of string Newton-Cartan geometry, following closely \cite{Bergshoeff:2019pij}. 

Let $\CT_p$ be the tangent space attached to a point $p$ in the spacetime $\CM$\,. We decompose $\CT_p$ into a longitudinal sector with an index $A = 0, 1$ and a transverse sector with an index $A' = 2, \cdots, d-1$\,. We introduce a longitudinal Vielbein field $\tau_\mu{}^A$ and a transverse Vielbein field $E_\mu{}^{A'}$. The invertibility conditions are
\begin{subequations}\label{eq:orthogandinvert}
\begin{align}
    \tau^\mu{}_{\!A} \, \tau_\mu{}^B & = \delta^B_A\,,
        &%
    \tau_\mu{}^A \, \tau^\nu{}_{\!A} + E_\mu{}^{A'} \, E^\nu{}_{\!A'} & = \delta^\nu_\mu\,, \\[2pt]
    E^\mu{}_{\!A'} \, E_\mu{}^{B'} & = \delta^{B'}_{A'}\,,
        &%
    \tau^\mu{}_{\!A} \, E_\mu{}^{A'} = E^\mu{}_{\!A'} \, \tau_\mu{}^A & = 0\,.
\end{align}
\end{subequations}
In addition, there is an additional gauge field $m_\mu{}^A$, associated with a noncentral extension in the string Newton-Cartan algebra. 

The string Newton-Cartan geometry realizes the string Newton-Cartan algebra \cite{Bergshoeff:2019pij} as a gauge symmetry acting on the target space that we will define momentarily. This string Newton-Cartan algebra is a finite subalgebra of the extended Galilean symmetry algebra discussed in \S\ref{sec:symmetries} for the flat spacetime free action in \eqref{eq:Snonrel} in the absence of a boundary. Turning on an arbitrary closed string background breaks the infinite-dimensional extended Galilean symmetry algebra to the finite-dimensional string Newton-Cartan symmetry algebra,\,\footnote{The string Newton-Cartan algebra can be further extended by including a longitudinal dilatation generator, similarly as in \eqref{eq:do} for the Bargmann algebra. See \cite{Bergshoeff:2019pij} for discussion on this dilatational generator.} whose generators consist of a longitudinal translation $H_{A}$\,, a transverse translation $P_{A'}$\,, a longitudinal Lorentz rotation $M$\,, a string Galilean boost $G_{AA'}$\,, a transverse rotation $J_{A'B'}$, and noncentral extensions $Z_A$ and $Z_{AB}$\,, with $\eta^{AB} Z_{AB} = 0$ and $\eta^{AB}$ the Minkowski metric. The string Newton-Cartan gauge fields transform under the string Newton-Cartan symmetry as \cite{Harmark:2018cdl, Bergshoeff:2019pij}
\begin{subequations} \label{eq:sNCtrnsf}
\begin{align}
    \delta \tau_\mu{}^A & = \Lambda \, \epsilon^A{}_B \, \tau_\mu{}^B\,, 
        \qquad
    \delta E_\mu{}^{A'} = - \Lambda_{A}{}^{A'} \tau_\mu{}^A + \Lambda^{A'}{}_{B'} E_\mu{}^{B'}\,, \\[2pt]
    \delta m_\mu{}^A & = D_\mu \sigma^A + \Lambda \, \epsilon^A{}_{B} \, m_\mu{}^B + \Lambda^{AA'} E_{\mu}{}^{A'} - \tau_\mu{}^B \, \sigma^A{}_B\,,
\end{align}
\end{subequations}
where $\sigma^{AB}$ is traceless with $\sigma^{AB} \eta^{}_{AB} = 0$ and $D_\mu \sigma^A = \p_\mu \sigma^A - \epsilon^A{}_B \, \sigma^B \, \Omega_\mu$ with $\Omega_\mu$ the spin connection associated with the longitudinal Lorentz rotation. The Levi-Civita symbol $\epsilon_{AB}$ is defined by $\epsilon_{01} = - \epsilon_{10} = 1$\,, and the $A$ index can be raised by the Minkowskian metric $\eta^{AB}$\,. Here, the Lie group parameter $\Lambda$ is associated with $M$, $\Lambda^{AA'}$ is associated with $G_{AA'}$, $\Lambda^{A'B'}$ is associated with $J_{A'B'}$, $\sigma^A$ is associated with $Z_A$, and $\sigma^{AB}$ is associated with $Z_{AB}$\,. Here, we omitted the symmetry transformations under translations and, instead, require that all gauge fields transform as covariant vectors under diffeomorphisms. These generators satisfy the string Newton-Cartan algebra with the following Lie brackets:
\begin{subequations} \label{eq:originalalgebra}
\begin{align}
    [H_A, G_{BA'}]  & = \eta^{}_{AB} P_{A'}\,,
        &%
    [H_A, Z]  & = \epsilon_{A}{}^{B} Z_B \,, \\[2pt]
    [G_{AA'}, P_{B'}] & = \delta_{A'\!B'} Z_A\,,
        &%
    [H_A, Z_{BC}]  & = 2 \, \eta_{AC} Z_B - \eta_{BC} Z_A\,, \\[2pt]
        & & %
    [G_{AA'}, G_{BB'}] & = \delta_{A'\!B'} Z_{[AB]}\,. 
\end{align} 
\end{subequations}
Here, we omitted the Lie brackets that involve the longitudinal and transverse rotational generators $M_{AB} = M \, \epsilon_{AB}$ and $J_{A'B'}$\,, which act on the $A$ and $A'$ indices in a standard way, for example, as in \eqref{eq:Bargmann}. 

Next, we consider the nonlinear sigma model in a string Newton-Cartan background. For now, we set the worldsheet to be flat and thus omit the dilaton field. We will return to the dilaton contribution at the end in \S\ref{sec:dilaton}. The nonlinear sigma model of nonrelativistic closed string theory in an arbitrary string Newton-Cartan background with a Kalb-Ramond field $B_{\mu\nu}$ is 
\begin{align} \label{eq:Sclosed}
    S_\text{closed} & = \frac{1}{4\pi\alpha'} \int d^2 \sigma \Bigl( \p_\alpha X^\mu \, \p^\alpha X^\nu H_{\mu\nu} + \lambda \, \overline{\p} X^\mu \, \tau_\mu + \overline{\lambda} \, \p X^\mu \, \overline{\tau}_\mu - i \, \epsilon^{\alpha\beta} \, \p_\alpha X^\mu \, \p_\beta X^\nu B_{\mu\nu} \Bigr)\,, \notag\\[2pt]
    H_{\mu\nu} & = E_\mu{}^{A'} E_\nu{}^{A'} + \lr \tau_\mu{}^A m_\nu{}^{B} + \tau_\nu{}^A m_\mu{}^{B} \rr \eta_{AB}\,,
\end{align}
where $\p = \p_\sigma - i \p_\tau$ and $\overline{\p} = \p_\sigma + i \p_\tau$ and the Levi-Civita symbol $\epsilon^{\alpha\beta}$ is defined by $\epsilon^{\tau\sigma} = - \epsilon^{\sigma\tau} = 1$\,. 
The $B$-field transforms under an $U(1)$ gauge symmetry 
\be \label{eq:gaugeB}
	\delta_\epsilon B_{\mu\nu} = \p_\mu \epsilon_\nu - \p_\nu \epsilon_\mu\,.
\ee
The two-tensor $H_{\mu\nu}$ is invariant under the string Galilean boost but not the $Z_A$ extension. For $S_\text{closed}$ to be invariant under the $Z_A$ symmetry that acts on $m_\mu{}^A$ as in \eqref{eq:sNCtrnsf}, it is required that a hypersurface orthogonality condition is satisfied,
\be \label{eq:hsoc}
    D_{[\mu} \, \tau_{\nu]}{}^A = 0\,.
\ee
With the condition \eqref{eq:hsoc} taken into account, the action \eqref{eq:Sclosed} is invariant under the gauge transformations \eqref{eq:sNCtrnsf} and \eqref{eq:gaugeB}. The $Z_A$ gauge symmetry we consider here prohibits the $\lambda \overline{\lambda}$ operator from being generated at the quantum level in the string action \eqref{eq:Sclosed}. Turning on such a $\lambda \overline{\lambda}$ term would drive the nonlinear sigma model towards the one that describes relativistic string theory. See \cite{Gomis:2019zyu, Yan:2019xsf} for details.\,\footnote{Note that $B_{\mu\nu}$ is taken to be invariant under the $Z_A$ gauge transformation here. In \cite{Harmark:2019upf}, a different spacetime gauge symmetry group is proposed, in which there is a different $Z_A$ symmetry transformation that acts nontrivially on both $m_\mu{}^A$ and $B_{\mu\nu}$\,, and the condition \eqref{eq:hsoc} is not imposed classically. In this latter case, a $\lambda \overline{\lambda}$ term in the sigma model will in general be generated by quantum corrections \cite{Gomis:2019zyu}. 
However, it is possible to relax the condition \eqref{eq:hsoc} to be $E^\mu{}_{\!A'} E^\nu{}_{\!B'} D_{[\mu} \tau_{\nu]}{}^A = 0$ without generating a $\lambda \overline{\lambda}$ operator at all loops in the two-dimensional worldsheet sigma model (which can be shown, for example, by using the method in \cite{Yan:2019xsf}). The condition $E^\mu{}_{\!A'} E^\nu{}_{\!B'} D_{[\mu} \tau_{\nu]}{}^A = 0$ is related to the twistless torsion conditions explored in \cite{Gallegos:2019icg}. There is no known symmetry reasoning for this type of twistless torsion conditions to hold from the worldsheet perspective.\label{fn:torsion}} 

Sometimes it is useful to rewrite the string action using the field redefinitions \cite{{Bergshoeff:2018yvt}}
\begin{subequations} \label{eq:fredef}
\begin{align}
    \lambda' = C^{-1} \! \lr \lambda - \p X^\mu \, \overline{C}_\mu \rr\,,
        \qquad
    \overline{\lambda}' = \overline{C}^{-1} \! \lr \overline{\lambda} - \overline{\p} X^\mu \, C_\mu \rr\,,
\end{align}
and
\begin{align}
    \tau'_\mu & = C \, \tau_\mu\,,
        \qquad
    H'_{\mu\nu} = H_{\mu\nu}- \lr C_\mu{}^A \, \tau_\nu{}^A + C_\nu{}^A \, \tau_\mu{}^B \rr \eta_{AB}\,, \\[2pt]
    \overline{\tau}_\mu & = \overline{C} \, \overline{\tau}_\mu\,,
        \qquad
    B'_{\mu\nu} = B_{\mu\nu} + \lr C_\mu{}^A \, \tau_\nu{}^B - C_\nu{}^A \, \tau_\mu{}^B \rr \epsilon_{AB}\,.
\end{align}
\end{subequations}
This can be thought of as a Stueckelberg symmetry in \eqref{eq:Sclosed}.
Here, $C_\mu = C_\mu{}^0 + C_\mu{}^1$ and $\overline{C}_\mu = C_\mu{}^0 - C_\nu{}^1$. The parameters $C$, $\overline{C}$ and $C_\mu{}^A$ are arbitrary functions of $X^\mu$ satisfying the condition $E^\mu{}_{\!A'} \, \p_\mu \! \lr C \overline{C} \rr=0$.
The dilaton field $\Phi$\,, when included, will also receive a redefinition,
\be \label{eq:dilatonshift}
    \Phi' = \Phi + \tfrac{1}{2} \ln \lr \overline{C} C \rr. 
\ee

\subsection{Dirichlet Sigma Model for Nonrelativistic Open Strings} \label{sec:Dsmnos}

We now consider nonrelativistic string theory defined by the worldsheet action \eqref{eq:Sclosed}  with a boundary $\p\Sigma$ at $\sigma = 0$ on the worldsheet $\Sigma$ and the boundary conditions corresponding to  a D$(d-2)$-brane,   described by a codimension-one submanifold $\CN$ embedded in the target space $\CM$\,. In a curved spacetime, we take
\be \label{eq:Dbdrycond}
    X^\mu \Big|_{\sigma = 0} = f^\mu (Y^i)\,,
\ee
where $Y^i$\,, $i = 0, 1, \cdots, d-2$ are coordinates in the submanifold $\CN$\,. The function $f^\mu$ in \eqref{eq:Dbdrycond} describes how the D$(d-2)$-brane is embedded in the $d$-dimensional spacetime. The $d$ conditions in \eqref{eq:Dbdrycond} imply that
\be
    \delta X^\mu \Big|_{\sigma = 0} = \delta Y^i \, \p_i f^\mu (Y)\,,
\ee
i.e., $\delta X^\mu \big|_{\p\Sigma}$ is tangent to $\CN$\,. The boundary condition \eqref{eq:Dbdrycond} generalizes the Dirichlet boundary condition \eqref{eq:flatBC1} in flat spacetime. 

\vspace{3mm}

\noindent \textbf{Unbroken Phase}

\vspace{3mm}

The D-brane submanifold $\CN$ spontaneously breaks the string Newton-Cartan symmetry group generated by \eqref{eq:originalalgebra} to the Bargmann symmetry group generated by \eqref{eq:Bargmann}. This theory can be studied in the unbroken and broken phase. The Nambu-Goldstone boson associated with this spontaneous symmetry breaking is the massless mode associated with perturbing the shape of the brane, which can be thought of as part of $f^\mu$ by allowing the brane to fluctuate. This defines the unbroken phase of the theory. 

In the unbroken phase, we introduce the background fields for open strings on the D-brane by exponentiating the vertex operator in \eqref{eq:Vbdry}, which modifies the closed string action \eqref{eq:Sclosed} to be
\begin{align} \label{eq:unbroken}
    S & = \frac{1}{4\pi\alpha'} \int_\Sigma d^2 \sigma \Bigl( \p_\alpha X^\mu \, \p^\alpha X^\nu H_{\mu\nu} + \lambda \, \overline{\p} X^\mu \, \tau_\mu + \overline{\lambda} \, \p X^\mu \, \overline{\tau}_\mu - i \, \epsilon^{\alpha\beta} \, \p_\alpha X^\mu \, \p_\beta X^\nu B_{\mu\nu} \Bigr) \notag \\[2pt]
    & \quad + \frac{1}{2\pi\alpha'} \int_{\p\Sigma} d\tau \, \Bigl[ \tfrac{1}{2} N(Y) \lr \lambda - \overline{\lambda} \rr + i \, A_i (Y) \, \p_\tau Y^i \Bigr]\,,
\end{align}
The definitions of the boundary background fields $N$ and $A_i$ here are the same as in flat spacetime up to a rescaling factor $2\pi\alpha'$\,. 
Varying $\lambda$ and $\overline{\lambda}$ on the boundary sets 
\be \label{eq:N=0}
    N(Y) = 0\,.
\ee
Furthermore, using \eqref{eq:Dbdrycond} and varying the total action with respect to the coordinates $Y^i$ on the brane, we find $d-1$ boundary conditions that generalize the Neumann boundary conditions \eqref{eq:flatBC4} and \eqref{eq:flatBC2} in flat spacetime, 
\be \label{eq:unbrokenphase}
    \p_i f^\mu \, H_{\mu\nu} \, \p_\sigma X^\nu + i \CF_{ij} \, \p_\tau X^j + \tfrac{1}{2} \, \p_i f^\mu \! \lr \lambda \, \tau_\mu + \overline{\lambda} \, \overline{\tau}_\mu \rr = 0\,,
\ee
where $\CF_{ij} \equiv b_{ij} + F_{ij}$\,, with $b_{ij} \equiv \p_i f^\mu \, B_{\mu\nu} \, \p_j f^\nu$ and $F_{ij} \equiv \p_i A_j - \p_j A_i$\,.  
Finally, the Lagrange multipliers $\lambda$ and $\overline{\lambda}$ impose the bulk equations of motion 
\be
    \overline{\p} X^\mu \, \tau_\mu = 0\,,
        \qquad
    \p X^\mu \, \overline{\tau}_\mu = 0\,,
\ee
which upon restricting  to the  boundary give the   boundary conditions
\begin{align} \label{eq:ahcc}
    \p_\sigma X^\mu \, \tau_\mu{}^A = i \, \epsilon^A{}_B \, \p_\tau Y^i \, \p_i f^\mu \, \tau_\mu{}^B\,.
\end{align}

In the unbroken phase, we choose to expand around the reference configuration defined by the boundary conditions \eqref{eq:N=0}--\eqref{eq:ahcc} with $N = 0$\,, and $N$ decouples from any further calculation on the beta-functions. However, there will still be counterterms generated for $N$, which will give rise to a nontrivial beta function $\beta (N)$\,. This is similar to the situation in relativistic open string theory \cite{Leigh:1989jq}.

The full string Newton-Cartan symmetry in \eqref{eq:sNCtrnsf} is preserved by the boundary conditions \eqref{eq:N=0}--\eqref{eq:ahcc} in this unbroken phase, provided that $\lambda$\,, $\overline{\lambda}$ transform on the brane as \cite{Bergshoeff:2019pij}
\begin{subequations} \label{eq:deltall}
\begin{align}
    \delta \lambda \, \big|_{\sigma = 0} & = \Lambda \lambda + \p X^\mu \ls D_\mu (\sigma^0 - \sigma^1) + \tau_\mu (\sigma^{00} - \sigma^{(01)}) \rs, \\[2pt]
    \delta \overline{\lambda} \, \big|_{\sigma = 0} & = - \Lambda \overline{\lambda} + \overline{\p} X^\mu \ls D_\mu (\sigma^0 + \sigma^1) + \overline{\tau}_\mu (\sigma^{00} + \sigma^{(01)}) \rs,
\end{align}
\end{subequations}
and that $A_i$ transforms as
\be \label{eq:deltaAi}
    \delta A_i = - \epsilon_{AB} \, \sigma^A \, \p_i f^\mu \, \tau_\mu{}^B\,,
\ee
The Stueckelberg symmetry in \eqref{eq:fredef}, with a trivial action on $A_i$\,,  is also preserved by the boundary conditions \eqref{eq:N=0}--\eqref{eq:ahcc}.

\vspace{3mm}

\noindent \textbf{Broken Phase}

\vspace{3mm}

Next, we consider the uniformly broken phase where the vacuum expectation value for $f^\mu$ is
\be
    \langle f^\mu (Y^i) \rangle = f^\mu_0 (Y^i)\,,
\ee
with $f^\mu_0$ a fixed embedding function that contains no massless excitations. Define a coordinate system $X^\mu = (y, Y^i)$ adapted to the submanifold specified by the embedding function $f_0^\mu (Y^i)$\,, with
\be
    f^y_0 = y_0\,,
        \qquad
    f^i_0 = Y^i,
\ee
where $y_0$ is a fixed location in the $y$ direction that determines where the brane is. We only need this set of adapted coordinates to be defined in a neighborhood of the submanifold. In nonrelativistic open string theory, we need to take the boundary values
\be \label{eq:tauEtaub}
    \tau_y{}^0 = E_y{}^{A'} = 0\,,
        \qquad
    \tau_y{}^1 \neq 0\,,
\ee
which naturally generalize the requirement that the submanifold is transverse to the $X^1$-direction in flat spacetime. 
It is convenient to rescale $\lambda$ and $\overline{\lambda}$ in \eqref{eq:unbroken} to normalize $\tau_y{}^1 = 1$ on the boundary. Applying these prescriptions to the invertibility conditions in \eqref{eq:orthogandinvert}, we find on the boundary that
\be \label{eq:pretE}
    \tau^y{}_1 = 1\,,
        \qquad
    \tau^i{}_1 = 0\,,
        \qquad
    \tau^y{}_ 0 = - \tau^i{}_0 \, \tau_i{}^1,
        \qquad
    E^y{}_{A'} = - E^i{}_{A'} \, \tau_i{}^1,
\ee
and
\begin{subequations} \label{eq:invsubm}
\begin{align}
    \tau^i{}_{0} \, \tau_i{}^0 & = 1\,,
        &%
    \tau_i{}^0 \, \tau^j{}_{0} + E_i{}^{A'} E^j{}_{\!A'} & = \delta^j_i\,, \\[2pt]
    E^i{}_{\!A'} \, E_i{}^{B'} & = \delta^{B'}_{A'}\,,
        &%
    \tau^i{}_{\!0} \, E_i{}^{A'} = E^i{}_{\!A'} \, \tau_i{}^0 & = 0\,.
\end{align}
\end{subequations}

Now, we consider the massless excitation that perturbs around the vacuum expectation value $f^\mu_0 (Y^i)$\,. We parametrize $f^\mu (Y^i)$ as
\be \label{eq:fyfi}
    f^y (Y^i) = y^{}_0 + \pi (Y^i)\,,
        \qquad
    f^i (Y^i) = Y^i,
\ee
where $\pi (Y^i)$ is the Nambu-Goldstone boson that perturbs around the brane. In nonrelativistic open string theory, it is possible to take a change of variables that is discontinuous on the boundary, under which $\pi (Y^i)$ contributes a $(\lambda-\overline{\lambda})$-term to the boundary action in \eqref{eq:unbroken} and thus gives rise to a nonzero $N = \pi$\,.\footnote{See similar discussions for relativistic open string theory in \cite{Leigh:1989jq, Kummer:2000ae}. Also see \cite{Tdualopenstring} for further details.}

The chosen vacuum expectation value of $f^\mu$ preserves the time translation generated by $H_0$\,, the transverse spatial translation generated by $P_{A'}$\,, the Galilean boost generated by $G_{0A'}$\,, and the transverse rotation generated by $J_{A'B'}$. These generators form the Bargmann algebra in \eqref{eq:Bargmann}.\footnote{The central charge $Z$ will be generated by commuting $P_{A'}$ and $G_{0A'}$\,.} In contrast, the translational symmetry generated by $H_1$\,, the longitudinal Lorentz rotation generated by $M$, and the part of the string Galilean boost symmetry generated by $G_{1A'}$ are broken on the boundary. This can be seen by requiring the transformations in \eqref{eq:sNCtrnsf} (with translations included) to preserve the boundary values in \eqref{eq:tauEtaub}.
 
\vspace{1mm}

In the following, we will proceed with the calculation of the beta-functions for the Dirichlet nonlinear sigma model \eqref{eq:unbroken}. We will mostly focus on the unbroken phase formalism, where the calculation is simpler. The broken phase formalism will be useful later when we vary the worldvolume DBI-like action to derive the equations of motion on the brane.

\subsection{Covariant Background Field Method}

In the rest of the section, we compute the beta-functions of the coupling $A_i$ and $N$ in the 2d QFT \eqref{eq:unbroken} around the classical configuration $N (Y) = 0$\,. We start with rewriting \eqref{eq:unbroken} by using the field redefinition \eqref{eq:fredef} with $A'_i = A_i$\,, $C = \overline{C} = 1$ and $C_\mu{}^A = m_\mu{}^A$\,, which gives rise to the following equivalent action: 
\begin{align} \label{eq:unbrokenra}
    S & = \frac{1}{4\pi\alpha'} \int_\Sigma d^2 \sigma \Bigl( \p_\alpha X^\mu \, \p^\alpha X^\nu E_{\mu\nu} + \lambda \, \overline{\p} X^\mu \, \tau_\mu + \overline{\lambda} \, \p X^\mu \, \overline{\tau}_\mu - i \, \epsilon^{\alpha\beta} \, \p_\alpha X^\mu \, \p_\beta X^\nu \mathscr{B}_{\mu\nu} \Bigr) \notag \\[2pt]
    & \quad + \frac{1}{2\pi\alpha'} \int_{\p\Sigma} d\tau \Bigl[ \tfrac{1}{2} \, N(Y) \lr \lambda - \overline{\lambda} \rr + i \, A_i (Y) \, \p_\tau Y^i \Bigr]\,,
\end{align}
where
\be
    E_{\mu\nu} = E_\mu{}^{A'} E_\nu{}^{A'},
        \qquad
    \mathscr{B}_{\mu\nu} = B_{\mu\nu} + \lr m_\mu{}^A \, \tau_\nu{}^B - m_\nu{}^A \, \tau_\mu{}^B \rr \epsilon_{AB}\,.
\ee
In this way of rewriting, $E_{\mu\nu}$ is invariant under the $Z_A$ symmetry but not invariant under the boost symmetry. This action is supplemented with the classical boundary conditions \eqref{eq:N=0}--\eqref{eq:ahcc}, which after the field redefinitions become
\begin{subequations} \label{eq:bdrycondun}
\begin{align}
    N = 0\,,  
        \qquad
    \tau_\mu{}^A \, \p_\sigma X^\mu & = i \, \epsilon^A{}_B \, \p_\tau Y^i \, t_i{}^B\,, \\[2pt]
    E_\mu{}^{A'} \p_\sigma X^\mu & = - e^i{}_{\!A'} \Bigl[ i \, \mathscr{F}_{ij} \, \p_\tau Y^j + \tfrac{1}{2} \lr \lambda - \overline{\lambda} \rr t_i{}^1 \Bigr], \\[2pt]
    \tfrac{1}{2} \lr \lambda + \overline{\lambda} \rr & = - t^{\,i}{}_{\,0} \, \Bigl[ i \, \mathscr{F}_{ij} \, \p_\tau Y^j + \tfrac{1}{2} \lr \lambda - \overline{\lambda} \rr t_i{}^1 \Bigr].
\end{align}
\end{subequations}
Here, we defined the projections
\be \label{eq:defetF}
    t_i{}^A = \p_i f^\mu \, \tau_\mu{}^A\,,
        \qquad
    e_i{}^{A'} = \p_i f^\mu \, E_\mu{}^{A'}\,,
        \qquad
    \mathscr{F}_{ij} = \p_i f^\mu \, \mathscr{B}_{\mu\nu} \, \p_j f^\nu + F_{ij}\,,
\ee
where $F_{ij} = \p_i A_j - \p_j A_i$\,.
The inverse Vielbeine fields $e^i{}_{A'}$ and $t^i{}_0$ are defined via the invertibility conditions on the brane,
\begin{subequations} \label{eq:etinv}
\begin{align}
    t^i{}_{0} \, t_i{}^0 & = 1\,,
        &%
    t_i{}^0 \, t^j{}_{0} + e_i{}^{A'} e^j{}_{\!A'} & = \delta^j_i\,, \\[2pt]
    e^i{}_{\!A'} \, e_i{}^{B'} & = \delta^{B'}_{A'}\,,
        &%
    t^i{}_{\!0} \, e_i{}^{A'} = e^i{}_{\!A'} \, t_i{}^0 & = 0\,.
\end{align}
\end{subequations}
Comparing with \eqref{eq:orthogandinvert}, we find the consistency conditions
\be \label{eq:tauraisemut}
   \tau^\mu{}_0 = \p_i f^\mu \, t^i{}_0\,,
        \qquad
    E^\mu{}_{A'} = \p_i f^\mu \, e^i{}_{A'}\,.
\ee

Next, we expand the sigma model \eqref{eq:unbrokenra} with respect to quantum fluctuations using the background field method, around a covariant background that satisfies the classical configurations \eqref{eq:N=0}--\eqref{eq:ahcc}. 

Consider a sufficiently small neighborhood $\CO_\CM$ of a point $X_0^\mu$ in the target space $\CM$\,. For an arbitrary point $X^\mu$ in $\CO_\CM$\,, there exists a unique geodesic in $\CM$ interpolating between $X_0^\mu$ and $X^\mu$, parametrized by $X_u^\mu$\,, with an affine parameter $u \in [0, 1]$, such that 
\be
    \frac{d^2 X_u^\mu}{du^2} + \Gamma^\mu{}_{\rho\sigma} \frac{d X^\rho_u}{du} \frac{d X^\sigma_u}{du} = 0\,.
\ee
We require $X_{u=0} = X_0^\mu$ and $X_{u=1} = X^\mu$\,. Here, $\Gamma^\mu{}_{\rho\sigma}$ is the Christoffel symbol in string Newton-Cartan geometry.\footnote{The explicit form of $\Gamma^\mu{}_{\rho\sigma}$ is given in \cite{Andringa:2012uz} but with $m_\mu{}^A$ set to zero, since the dependence on $m_\mu{}^A$ has been relocated to be in $\mathscr{B}_{\mu\nu}$\,.} 
. Define the covariant quantum fluctuation
\be
    \xi^\mu = \frac{d X^\mu_u}{du} \bigg|_{u=0}\,,
\ee
which is the tangent vector to the geodesic at $u = 0$\,.
Similarly, consider a sufficiently small neighborhood $\CO_\CN$ of a point $Y_0^i$ in the submanifold $\CN$. For an arbitrary point $Y^i$ in $\CO_\CN$\,, there exists a unique geodesic in $\CN$ interpolating between $Y_0^i$ and $Y^i$\,, parametrized by $Y_v^i$\,, with an affine parameter $v \in [0, 1]$, such that 
\be
    \frac{d^2 Y_v^i}{dv^2} + \Gamma^i{}_{jk} \frac{d Y^j_v}{dv} \frac{d Y^k_v}{dv} = 0\,.
\ee
We require $Y_{v=0} = Y_0^i$ and $Y_{v=1} = Y^i$\,. Here, $\Gamma^i{}_{jk}$ is the Christoffel symbol in Newton-Cartan geometry on the brane. Define the covariant quantum fluctuation
\be
    \zeta^i = \frac{d Y^i_v}{dv} \bigg|_{v=0}\,,
\ee
which is the tangent vector to the geodesic at $v = 0$\,. The tangent vector $\zeta^i (Y)$ on the brane submanifold $\CN$ is related to the tangent vector $\xi^\mu (X)$ in $\CM$ as follows \cite{Leigh:1989jq}:
\be \label{eq:xitozeta}
    \xi^\mu \big|_{\sigma = 0} = \zeta^i \, \p_i f^\mu + \tfrac{1}{2} \, \zeta^i \, \zeta^j \, K^\mu{}_{ij} + \cdots\,,
\ee
where $K^\mu{}_{ij}$ is the extrinsic curvature of the brane submanifold, defined by
\be
    K^\mu_{ij} = \nabla_{\!i} \, \nabla_{\!j} \, f^\mu + \Gamma^\mu{}_{\rho\sigma} \, \p_{i} f^\rho \, \p_{j} f^\sigma\,.
\ee
Here, $\nabla_{\!i}$ is a Newton-Cartan covariant derivative in the submanifold $\CN$ and $\Gamma^\mu{}_{\rho\sigma}$ is the Christoffel symbol in string Newton-Cartan geometry, with
\be \label{eq:Christoffel}
    \Gamma^\mu{}_{\rho\sigma} \, \big|_{\sigma = 0} = \tau^\mu{}_{\!A} \, \p_{(\rho} \tau_{\sigma)}{}^A + E^\mu{}_{\!A'} \! \ls \p^{}_{(\rho} E^{}_{\sigma)}{}^{A'} - \Omega_{(\rho}{}^{A'B'} E^{}_{\sigma)}{}^{B'} - \Omega_{(\rho}{}^{0A'} \tau^{}_{\sigma)}{}^0 \rs\!,
\ee
where $\Omega_\mu{}^{AA'}$ is the spin connection associated with the string Galilean boost and $\Omega_\mu{}^{A'B'}$ is the spin connection associated with the transverse rotation. We used that $\Omega_\mu{}^{AB} \big|_{\sigma = 0} = \Omega_\mu{}^{1A'} \big|_{\sigma = 0} = 0$ since they are associated with spontaneously broken symmetries, where $\Omega_\mu{}^{AB}$ is the spin connection associated with the longitudinal Lorentz boost. 
In the adapted coordinates \eqref{eq:fyfi}, we also have the boundary conditions $\Omega_y{}^{0A'} \big|_{\sigma = 0} = \Omega_y{}^{A'B'} \big|_{\sigma = 0} = 0$\,. It then follows that, on the boundary, 
\be \label{eq:tK}
    E_\mu{}^{A'} K^\mu{}_{ij} = \tau_\mu{}^0 K^\mu{}_{ij} = 0\,,
        \qquad
    \tau_\mu{}^1 K^\mu{}_{ij} = \nabla_{\!i} \, t_j{}^1 = \nabla_{\!j} \, t_i{}^1\,.
\ee

A covariant expansion with respect to $\xi^\mu (X)$ for the closed string action, i.e. the bulk part in \eqref{eq:unbrokenra}, has been put forward in \cite{Yan:2019xsf}, where $\lambda$ and $\overline{\lambda}$ are also split into a classical and quantum part as
\be
    \lambda = \lambda_0 + \rho\,,
        \qquad
    \overline{\lambda} = \overline{\lambda}_0 + \rho\,. 
\ee
By incorporating  a covariant expansion with respect to $\zeta^i (Y)$ for the boundary action in \eqref{eq:unbrokenra}, in a similar way as in \cite{Leigh:1989jq}, we find the expanded action to be
\begin{align} \label{eq:covexp}
    S [\lambda\,, \overline{\lambda}\,, X\,, Y] = S^{(0)} + S^{(1)} + S^{(2)} + O(\rho\,, \overline{\rho}\,, \xi\,, \zeta)^3,  
\end{align}
where $S^{(0)} = S [\lambda_0\,, \overline{\lambda}_0\,, X_0\,, Y_0]$ and
\begin{align}
    S^{(1)} & = \frac{1}{4\pi\alpha'} \! \int_\Sigma d^2 \sigma \, \Bigl\{ \xi^\rho \Bigl[ \p_\alpha X^\mu_0 \, \p^\alpha X^\nu_0 \nabla_{\!\rho} E_{\mu\nu} - 2 \, \nabla_{\!\alpha} \! \lr \p^\alpha \! X_0^\mu \, E_{\rho\mu} \rr - i \, \epsilon^{\alpha\beta} \p_{\!\alpha} X^\mu_0 \, \p_\beta X^\nu_0 \, \mathscr{H}_{\mu\nu\rho} \notag \\[2pt]
    & \hspace{6.1cm} - \! \overline{\nabla} \! \lr \lambda_0 \, \tau_{\!\rho} \rr - \nabla \! \lr \overline{\lambda}_0 \, \overline{\tau}_{\!\rho} \rr \Bigr] + \rho \, \overline{\p} X^\mu_0 \, \tau_\mu + \overline{\rho} \, \p X^\mu_0 \, \overline{\tau}_\mu \Bigr\} \notag \\[2pt]
    & \quad + \frac{1}{2\pi\alpha'} \int_{\p\Sigma} d\tau \, \Bigl\{ \xi^\mu \! \ls \p_\sigma X^\nu_0 \, E_{\mu\nu}  + i \, \p_{\tau} Y^i_0 \, \mathscr{B}_{\mu i} + \tfrac{1}{2} \bigl( \lambda_0 \, \tau_\mu + \overline{\lambda}_0 \, \overline{\tau}_\mu \bigr) \rs + i \, \zeta^i \p_\tau Y^j_0 \, {F}_{ij} \Bigr\}\,, \notag \\[6pt]
    S^{(2)} & = \frac{1}{4\pi\alpha'} \int_\Sigma d^2 \sigma \, \Bigl\{ \nabla_{\!\alpha} \xi^\mu \, \nabla^\alpha \xi^\nu \, E_{\mu\nu} - \xi^\mu \! \ls \overline{\nabla} \! \lr \rho \, \tau_\mu \rr + \nabla \! \lr \overline{\rho} \, \overline{\tau}_{\!\mu} \rr \rs \notag \\
    & \hspace{2.7cm} + 2 \, \xi^\rho \, \nabla_{\!\alpha} \xi^\sigma \lr \p^\alpha X_0^\mu \, \nabla_{\![\rho} E_{\sigma]\mu} - \tfrac{i}{2} \, \epsilon^{\alpha\beta} \, \p_\beta X^\mu_0 \, \mathscr{H}_{\mu\rho\sigma} \rr \notag \\[6pt]
    & \hspace{2.7cm} + \tfrac{1}{2} \, \xi^\rho \xi^\sigma \, \p_\alpha X^\mu_0 \, \p_\beta X^\nu_0 \Big[ \delta^{\alpha\beta} \! \lr \nabla_{\!\rho} \nabla_{\!\sigma} E_{\mu\nu} - 2 \, \nabla_{\!\mu} \nabla_{\!\rho} E_{\sigma\nu} \rr - i \, \epsilon^{\alpha\beta} \, \nabla_{\!\rho} \mathscr{H}_{\sigma\mu\nu} \Bigr] \notag \\[4pt]
    & \hspace{2.7cm} 
    + \xi^\rho \xi^\sigma \Bigl[ \p_\alpha X^\mu_0 \, \p^\alpha X^\nu_0 \, E_{\mu \lambda} \, R^{\lambda}{}_{\rho\sigma \nu} 
    + \tfrac{1}{2} \bigl( \lambda_0 \, \overline{\p} X^\mu_0 \, \tau_\kappa + \overline{\lambda}_0 \, {\p} X^\mu_0 \, \overline{\tau}_\kappa \bigr) R^\kappa{}_{\rho\sigma \mu} \Bigr] \Bigr\} \notag \\[2pt]
    & \quad + \frac{1}{4\pi\alpha'} \int_{\p\Sigma} d\tau \, \Bigl[ i \, \xi^\mu \nabla_{\!\tau} \xi^\nu \mathscr{B}_{\mu\nu} + i \, \zeta^i \nabla_{\!\tau} \zeta^j F_{ij} + \rho \, \xi^\mu \, \tau_\mu + \overline{\rho} \, \xi^\mu \, \overline{\tau}_\mu \notag \\[2pt]
    & \hspace{5.5cm} + \zeta^i \zeta^j \lr \p_\sigma X^\rho_0 \, \p_i f^\mu \, \p_j f^\nu \, \nabla_{\!\mu} E_{\nu\rho} + i \, \p_\tau Y_0^k \, \nabla_{\!i} \mathscr{F}_{jk} \rr \Bigr]. \notag
\end{align}
It is understood that all the couplings are evaluated at $X = X_0$ and $Y = Y_0$\,. The background fields are required to satisfy the boundary conditions in \eqref{eq:bdrycondun} and therefore $N (Y_0) = 0$\,. Here, $\mathscr{F}_{ij}$ is defined in \eqref{eq:defetF} and 
\be
    \mathscr{H}_{\mu\nu\rho} = \p_\mu \mathscr{B}_{\nu\rho} + \p_\rho \mathscr{B}_{\mu\nu} + \p_\nu \mathscr{B}_{\rho\mu}\,.
\ee
The Riemann tensor is defined with respect to the Christoffel symbol $\Gamma^\mu{}_{\rho\sigma}$\,.   

To proceed to quantum calculations, it is convenient to change variables from $\xi^\mu$ to $(\xi^A, \xi^{A'})$ and from $\zeta^i$ to $\zeta^I = (\zeta^0, \zeta^{A'})$\,, with
\begin{align} \label{eq:nv}
    \xi^A = \tau_\mu{}^A \, \xi^\mu\,,
        \qquad
    \xi^{A'} = E_\mu{}^{A'} \xi^\mu\,,
        \qquad
    \zeta^0 = t_i{}^0 \, \zeta^i\,,
        \qquad
    \zeta^{A'} = e_i{}^{A'} \zeta^i\,.
\end{align}
In terms of these new variables, we have a simple propagator that is diagonalized. From \eqref{eq:covexp}, we find that the free part of the quadratic action is
\be \label{eq:freeaction}
    S_\text{free} = \frac{1}{4\pi\alpha'} \int_\Sigma d^2 \sigma \lr \p_\alpha \xi^{A'} \p^\alpha \xi^{A'} + \rho \, \overline{\p} \xi + \overline{\rho} \, \p \overline{\xi} \rr,
\ee
with $\xi = \xi^0 + \xi^1$ and $\overline{\xi} = \xi^0 - \xi^1$\,. 
At the zeroth order in quantum fluctuations, $\xi^A$, $\xi^{A'}$, $\rho$ and $\overline{\rho}$ satisfy the boundary conditions \eqref{eq:flatBC1}--\eqref{eq:flatBC5}\,, with
\be \label{eq:fluctuationsbconds}
    \p_\tau \xi^1 \, \big|_{\sigma = 0} = \p_\sigma \xi^0 \, \big|_{\sigma = 0} = \p_\sigma \xi^{A'} \, \big|_{\sigma=0} = \rho + \overline{\rho} \, \big|_{\sigma = 0} = 0\,,
        \qquad
    \p_\sigma \xi^1 + i \, \p_\tau \xi^0 \, \big|_{\sigma = 0} = 0\,.
\ee
It follows from \cite{Yan:2019xsf} that
\be
    \nabla_\alpha \xi^\mu \, \big|_{\sigma = 0} = \tau^\mu{}_{\!A} \, \p_\alpha \xi^A + E^\mu{}_{\!A'} \! \ls \p_\alpha \xi^{A'} - \bigl( \Omega_\nu{}^{0A'} \xi^{0} + \Omega_\nu{}^{A'B'} \xi^{B'} \big) \p_\alpha X^\nu_0 \rs\,.
\ee
There is ambiguity in identifying the covariant quantum fluctuation in ${(\rho - \overline{\rho}) \, \big|_{\sigma = 0}}$ that one should integrate out on the boundary. It will prove to be useful to introduce the following decomposition of $\rho - \overline{\rho}$ on the boundary:
\be \label{eq:rhozeta}
    \tfrac{1}{2} \lr \rho - \overline{\rho} \rr \big|_{\sigma = 0} = r - \tfrac{i}{2} \, \omega_{A'} \, \zeta^{A'} - \tfrac{i}{2} \, \p_\tau \bigl( \omega_{A'B'} \, \zeta^{A'} \zeta^{B'} \bigr) + O(\zeta^3)\,.
\ee
We will show that $\omega_{A'}$ and $\omega_{A'B'}$ are fixed by requiring that the one-loop effective action is gauge covariant.
Using \eqref{eq:xitozeta} that relates the bulk quantum field $\xi^\mu$ to the boundary quantum field $\zeta^i$\,, in terms of the new variables \eqref{eq:nv} which satisfy the free field boundary conditions \eqref{eq:fluctuationsbconds}, we read off the terms in \eqref{eq:covexp} that are quadratic in quantum fields as
\begin{align} \label{eq:S(2)}
    S_2 & = - \frac{i}{2\pi\alpha'} \int_\Sigma d^2\sigma \, \xi^I \p_\tau \xi^J \, \CW_{IJ} \notag \\[2pt]
    & \quad + \frac{1}{4\pi\alpha'} \int_{\p\Sigma} d\tau \Bigl( \zeta^I \, \zeta^J \, \CM_{IJ} + i \, \zeta^I \p_\tau \zeta^J \mathscr{F}_{IJ} + 2 \, r \, \zeta^I \, t_I{}^1 \Bigr),
\end{align}
where
\begin{align}
    \CW_{IJ} & = W_{IJ} + i \, \Omega_{\mu IJ} \, \p_\tau X^\mu_0\,, 
        \qquad
    \CM_{IJ} = M_{IJ} + M^\Omega_{IJ}\,,
\end{align} 
with
\begin{align*}
    M_{ij} & = i \, \p_\tau Y^k_0 \Bigr[ \nabla_{(i} \mathscr{F}_{j)k} + \lr K^\mu{}_{ij} \, \p_k f^\nu + K^\nu{}_{k(i} \, \p_{j)} f^\mu \rr \mathscr{B}_{\mu\nu} \Bigr] + \tfrac{1}{2} \lr \lambda - \overline{\lambda} \rr K^\mu{}_{ij} \, \tau_\mu{}^1, \\[2pt]
    M^\Omega_{00} & = - i \, \Omega_\tau{}^{0A'} \mathscr{F}_{0A'}\,, 
        \qquad\qquad\qquad\!\!%
    W_{\mu\nu} = \tfrac{1}{2} \, \mathscr{H}_{\mu\nu\rho} \, \p_\sigma X^\rho_0\,, \\[2pt]
    M^\Omega_{0A'} & = M^\Omega_{A'0} = - \tfrac{i}{2} \lr \Omega_{\tau}{}^{0B'} \mathscr{F}_{A'B'} - \Omega_\tau{}^{A'B'} \mathscr{F}_{0B'} + \omega_{A'} \, t_{0}{}^1 \rr\,, \\[2pt]
    M^\Omega_{\!A'B'} \! & = i \! \ls \Omega_{\tau (A'}{}^{C'} \mathscr{F}_{B')C'} - \, \p_\tau Y_0^i \, t_i{}^1 \! \lr E^\nu{}_{\!A'} \, \Omega_\nu{}^{0B'} - \omega_{A'B'} \rr
    - \, \omega^{}_{(A'} \, t_{B')}{}^1 \rs. 
\end{align*}
The boundary conditions in \eqref{eq:bdrycondun} have been applied to derive \eqref{eq:S(2)}. Note that $W_{IJ}$ is $W_{\mu\nu}$ projected by $\tau^\mu{}_0$ and $E^\mu{}_{\!A'}$\,. Moreover, $M_{IJ}, t_I{}^1$ and $\mathscr{F}_{IJ}$ are $M_{ij}, t_i{}^1$ and $\mathscr{F}_{ij}$ projected by $t^i{}_0$ and $e^i{}_{\!A'}$, respectively. We also defined $\Omega_\tau{}^{IJ} = \p_\tau Y^k_0 \, \p_k f^\mu \, \Omega_\mu{}^{IJ}$\,.
Here, we only kept terms that will give rise to nontrivial contributions to the one-loop boundary effective action. 
To derive \eqref{eq:S(2)}, we used the following identities:
\begin{subequations}
\begin{align}
    & \nabla_{\!\tau} \, \xi^\mu \, \big|_{\sigma=0} = \p_i f^\mu \, \nabla_{\!\tau} \zeta^i + \zeta^i K^\mu{}_{ij} \, \p_\tau Y^j_0 + O (\zeta^2)\,, \\[2pt]
    & \tau^\mu{}_0 \, \Omega_\mu{}^{0A'} = 0\,,
        \qquad\,\,\,\,
    E^\mu{}_{\!A'} \, \Omega_{\mu}{}^0{}_{B'} = E^\mu{}_{\!B'} \, \Omega_{\mu}{}^0{}_{A'}\,. 
\end{align}
\end{subequations}
In the next subsection, we compute the one-loop effective action by integrating out the quantum fluctuations in the path integral associated with the action \eqref{eq:S(2)}.

\subsection{Beta-Functions for Open String Couplings}

From the free action \eqref{eq:freeaction}, we read off the propagators, which are already given in \eqref{eq:XX} and \eqref{eq:lambdaX},
\begin{subequations} \label{eq:prop}
\begin{align}
    \langle \xi^{A'} (\tau\,, \sigma) \, \xi^{B'} (\tau'\,, \sigma') \rangle & = 2 \pi \alpha' \Big[ \Delta (\tau - \tau', \sigma - \sigma') + \Delta (\tau - \tau', \sigma + \sigma') \Bigr]\,, \\[2pt]
    \langle \rho (\tau\,, \sigma) \, \xi (\tau', \sigma') \rangle & = 4 \pi \alpha' \, \p \Delta (\tau - \tau', \sigma - \sigma')\,, \\[5pt]
    \langle \overline{\rho} (\tau\,, \sigma) \, \overline{\xi} (\tau', \sigma') \rangle & = 4 \pi \alpha' \, \overline{\p} \Delta (\tau - \tau', \sigma - \sigma')\,,
\end{align}
\end{subequations}
where we defined 
\begin{align}
    \Delta (\tau, \sigma) & = - \frac{1}{4\pi} \ln \bigl( \tau^2 + \sigma^2 \bigr) = \int \frac{d\omega \, dk}{(2\pi)^2} \, \frac{e^{ i ( \omega \, \tau + k \, \sigma)}}{\omega^2 + k^2} \,.
\end{align}
Since we are focusing on the boundary one-loop effective action, we are mostly interested in the boundary propagators with $\sigma = 0$\,. Because $\rho\,, \overline{\rho}$ and $\xi^A$ only appear as boundary fields in the contributions we are interested in, we can set $\rho (\tau, 0) = - \overline{\rho} (\tau, 0) \sim r (\tau)$ in \eqref{eq:prop}. It follows that the boundary-boundary propagators are
\begin{subequations}
\begin{align}
    \langle \xi^{A'} \! (\tau, 0) \, \xi^{B'} \! (\tau', 0) \rangle & = 4 \pi \alpha' \Delta (\tau - \tau', 0)\,, \\[2pt]
    \langle r (\tau) \, \xi^0 (\tau', 0) \rangle & = - 4 \pi i \, \alpha' \p_\tau \Delta (\tau - \tau', 0)\,, \\[2pt]
    \langle r (\tau) \, \xi^1 (\tau', 0) \rangle & = 0\,.
\end{align}
\end{subequations}
Introducing the index $\CA = (r\,, I)$\,, we write the boundary-boundary propagator as
\be \label{eq:propGab}
    \Delta^{\CA\CB} (\tau - \tau',0) = 
        2 \pi \alpha'%
    \begin{pmatrix}
        0 &\quad - i \, \p_{\tau} & \quad 0\\[2pt]
        i \, \p_{\tau} & \quad 0 &\quad 0\\[2pt]
    0 & \quad 0 & \quad \delta^{A' B'}
    \end{pmatrix}
    \Delta(\tau - \tau')\,,
\ee
where we defined 
\be \label{eq:Deltatau}
    \Delta (\tau) \equiv 2 \Delta (\tau, 0) = \int \frac{d\omega}{2\pi} \frac{e^{i \, \omega \, \tau}}{|\omega|} = - \frac{1}{\pi} \ln \left|\frac{\tau}{\tau^{}_\text{IR}}\right|,
\ee
with $\tau_\text{IR}$ an infrared regulator. Moreover, using a sharp-cutoff regularization in the frequency space, we find
\be
    \Delta (0) = \frac{1}{\pi} \log \lr \frac{\Lambda}{\mu} \rr,
\ee
with $\Lambda$ the ultraviolet cutoff and $\mu$ the infrared cutoff of the frequency.
We introduced a prefactor 2 in the definition \eqref{eq:Deltatau} of $\Delta (\tau)$\,, such that
\be
    \int d\tau'' \, \Delta^{-1} (\tau - \tau'') \, \Delta (\tau'' - \tau') = \delta (\tau - \tau')\,,
        \qquad
    \Delta^{-1} (\tau, \tau') \equiv \p_{\tau} \p_{\tau'} \Delta (\tau - \tau')\,.
\ee

From the quadratic action \eqref{eq:S(2)} that collects terms relevant to the one-loop boundary effective action, we read off the following Feynman rules for different vertices:
\begin{subequations}
\begin{align}
    V^\CW_{\!\CA\CB} (\tau, \sigma\,; \tau', \sigma') & = 
        \frac{i}{\pi \alpha'}
    \begin{pmatrix}
        0 & \quad 0 \\[2pt]
        0 & \quad \CW_{IJ} (\tau\,, \sigma) \, \p_\tau
    \end{pmatrix}
    \delta (\tau - \tau') \, \delta (\sigma - \sigma')\,, \\[2pt]
    V^{\mathscr{F}}_{\!\CA\CB} (\tau, \tau') & = - \frac{1}{2\pi\alpha'} 
    \begin{pmatrix}
        0 & \quad t^{}_{\!J}{}^1 \\[2pt]
        t_{I}{}^1 & \quad i \, \mathscr{F}_{IJ} (\tau) \, \p_\tau
    \end{pmatrix}
        \delta (\tau - \tau')\,, \\[2pt]
    V^\CM_{\!\CA\CB} (\tau, \tau') & = - \frac{1}{2\pi\alpha'} 
    \begin{pmatrix}
        0 & \quad 0 \\[2pt]
        0 & \quad \CM_{IJ}
    \end{pmatrix}
        \delta (\tau - \tau')\,.
\end{align}
\end{subequations}
The loop calculation is very similar to the procedure discussed in \cite{Leigh:1989jq, Abouelsaood:1986gd}. 
All the one-loop diagrams are collected below, 
\vspace{-5mm}
\begin{align} \label{eq:Fdiag1}
\Gamma^{\mathscr{\CF}}_n = 
\begin{minipage}{5cm}
\begin{tikzpicture}
    \draw [line width=0.55mm, black] (0,0) to (-4,0);
    \draw (-3.5,0) .. controls (-3.4,.45)  and (-2.5,.45) .. (-2.5,0);
    \filldraw [white] (-2.5,0) circle [radius=0.115];
    \draw (-2.5,0) circle [radius=0.115];
    \draw (-2.5,0) .. controls (-2.4,.45)  and (-1.5,.45) .. (-1.5,0);
    \filldraw [black] (-1.2,.15) circle [radius=0.02];
    \filldraw [black] (-1.05,.15) circle [radius=0.02];
    \filldraw [black] (-.9,.15) circle [radius=0.02];
    \draw (-3.5,0) .. controls (-3.5,1.65) and (-0.5,1.65) .. (-0.5,0);
    \filldraw [white] (-3.5,0) circle [radius=0.115];
    \draw (-3.5,0) circle [radius=0.115];
    \node at (-3.5,0) {\scalebox{0.85}{$\times$}};
    \node at (-3.5,-.4){$V^\mathscr{F}$};
    \filldraw [white] (-1.5,0) circle [radius=0.115];
    \draw (-1.5,0) circle [radius=0.115];
    \node at (-1.5,0) {\scalebox{0.85}{$\times$}};
    \filldraw [white] (-2.5,0) circle [radius=0.115];
    \draw (-2.5,0) circle [radius=0.115];
    \node at (-2.5,0) {\scalebox{0.85}{$\times$}};
    \node at (-2.5,-.4){$V^\mathscr{F}$};
    \filldraw [white] (-.5,0) circle [radius=0.115];
    \draw (-.5,0) circle [radius=0.115];
    \node at (-.5,0) {\scalebox{0.85}{$\times$}};
    \node at (-0.5,-.4){$V^\mathscr{F}$};
    \filldraw [black] (-1.3-.1,-.45) circle [radius=0.02];
    \filldraw [black] (-1.15-.1,-.45) circle [radius=0.02];
    \filldraw [black] (-1-.1,-.45) circle [radius=0.02];
    \filldraw [black] (-1.3-0.5-.1,-.45) circle [radius=0.02];
    \filldraw [black] (-1.15-0.5-.1,-.45) circle [radius=0.02];
    \filldraw [black] (-1-0.5-.1,-.45) circle [radius=0.02];
    \node at (.5,0) {\scalebox{0.8}{$\sigma \! = \! 0$}};
    \node at (-1.7,-.73) {\scalebox{0.75}{$\underbrace{\hspace{2.75cm}}$}};
    \node at (-1.7,-1) {\scalebox{0.8}{$n$}};
    \node at (-2,1.6) {$\phantom{\CW}$};
    \node at (0,-1.8cm) {$\phantom{ }$};
\end{tikzpicture}
\end{minipage}
\qquad\quad
\Gamma^\CM_n = 
\begin{minipage}{5cm}
\begin{tikzpicture}
    \draw [line width=0.55mm, black] (0,0) to (-4,0);
    \draw (-3.5,0) .. controls (-3.4,.45)  and (-2.5,.45) .. (-2.5,0);
    \filldraw [white] (-2.5,0) circle [radius=0.115];
    \draw (-2.5,0) circle [radius=0.115];
    \draw (-2.5,0) .. controls (-2.4,.45)  and (-1.5,.45) .. (-1.5,0);
    \filldraw [black] (-1.2,.15) circle [radius=0.02];
    \filldraw [black] (-1.05,.15) circle [radius=0.02];
    \filldraw [black] (-.9,.15) circle [radius=0.02];
    \draw (-3.5,0) .. controls (-3.5,1.65) and (-0.5,1.65) .. (-0.5,0);
    \filldraw [white] (-3.5,0) circle [radius=0.115];
    \draw (-3.5,0) circle [radius=0.115];
    \node at (-3.5,0) {\scalebox{0.85}{$\times$}};
    \node at (-3.5,-.4){$V^\mathcal{M}$};
    \filldraw [white] (-1.5,0) circle [radius=0.115];
    \draw (-1.5,0) circle [radius=0.115];
    \node at (-1.5,0) {\scalebox{0.85}{$\times$}};
    \filldraw [white] (-2.5,0) circle [radius=0.115];
    \draw (-2.5,0) circle [radius=0.115];
    \node at (-2.5,0) {\scalebox{0.85}{$\times$}};
    \node at (-2.5,-.4){$V^\mathscr{F}$};
    \filldraw [white] (-.5,0) circle [radius=0.115];
    \draw (-.5,0) circle [radius=0.115];
    \node at (-.5,0) {\scalebox{0.85}{$\times$}};
    \node at (-0.5,-.4){$V^\mathscr{F}$};
    \filldraw [black] (-1.3-.1,-.45) circle [radius=0.02];
    \filldraw [black] (-1.15-.1,-.45) circle [radius=0.02];
    \filldraw [black] (-1-.1,-.45) circle [radius=0.02];
    \filldraw [black] (-1.3-0.5-.1,-.45) circle [radius=0.02];
    \filldraw [black] (-1.15-0.5-.1,-.45) circle [radius=0.02];
    \filldraw [black] (-1-0.5-.1,-.45) circle [radius=0.02];
    \node at (.5,0) {\scalebox{0.8}{$\sigma \! = \! 0$}};
    \node at (-1.7,-.73) {\scalebox{0.75}{$\underbrace{\hspace{2.75cm}}$}};
    \node at (-1.7,-1) {\scalebox{0.8}{$n$}};
    \node at (-2,1.6) {$\phantom{\CW}$};
    \node at (0,-1.8cm) {$\phantom{ }$};
\end{tikzpicture}
\end{minipage}
\end{align}
\vspace{-1cm}\\
and
\vspace{-5mm}
\begin{align} \label{eq:Fdiag2}
\Gamma^\CW_n =
\begin{minipage}{5cm}
\begin{tikzpicture}
    \draw [line width=0.55mm, black] (0,0) to (-4,0);
    \draw (-3.5,0) .. controls (-3.4,.45)  and (-2.5,.45) .. (-2.5,0);
    \filldraw [white] (-2.5,0) circle [radius=0.115];
    \draw (-2.5,0) circle [radius=0.115];
    \draw (-2.5,0) .. controls (-2.4,.45)  and (-1.5,.45) .. (-1.5,0);
    \draw (-1.20,.15) circle [radius=0.02];
    \filldraw [black] (-1.2,.15) circle [radius=0.02];
    \draw (-1.05,.15) circle [radius=0.02];
    \filldraw [black] (-1.05,.15) circle [radius=0.02];
    \draw (-.9,.15) circle [radius=0.02];
    \filldraw [black] (-.9,.15) circle [radius=0.02];
    \draw (-3.5,0) .. controls (-3.5,1.65) and (-0.5,1.65) .. (-0.5,0);
    \filldraw [white] (-3.5,0) circle [radius=0.115];
    \draw (-3.5,0) circle [radius=0.115];
    \node at (-3.5,0) {\scalebox{0.85}{$\times$}};
    \node at (-3.5,-.4){$V^\mathscr{F}$};
    \filldraw [white] (-1.5,0) circle [radius=0.115];
    \draw (-1.5,0) circle [radius=0.115];
    \node at (-1.5,0) {\scalebox{0.85}{$\times$}};
    \filldraw [white] (-2.5,0) circle [radius=0.115];
    \draw (-2.5,0) circle [radius=0.115];
    \node at (-2.5,0) {\scalebox{0.85}{$\times$}};
    \node at (-2.5,-.4){$V^\mathscr{F}$};
    \filldraw [white] (-.5,0) circle [radius=0.115];
    \draw (-.5,0) circle [radius=0.115];
    \node at (-.5,0) {\scalebox{0.85}{$\times$}};
    \node at (-0.5,-.4){$V^\mathscr{F}$};
    \filldraw [black] (-1.3-.1,-.45) circle [radius=0.02];
    \filldraw [black] (-1.15-.1,-.45) circle [radius=0.02];
    \filldraw [black] (-1-.1,-.45) circle [radius=0.02];
    \filldraw [black] (-1.3-0.5-.1,-.45) circle [radius=0.02];
    \filldraw [black] (-1.15-0.5-.1,-.45) circle [radius=0.02];
    \filldraw [black] (-1-0.5-.1,-.45) circle [radius=0.02];
    \filldraw [white] (-2,1.25) circle [radius=0.115];
    \draw (-2,1.25) circle [radius=0.115];
    \node at (-2,1.25) {\scalebox{0.85}{$\times$}};
    \node at (-2,1.6) {$\CW$};
    \node at (.5,0) {\scalebox{0.8}{$\sigma \! = \! 0$}};
    \node at (-1.7,-.73) {\scalebox{0.75}{$\underbrace{\hspace{2.75cm}}$}};
    \node at (-1.7,-1) {\scalebox{0.8}{$n$}};
    \node at (0,-1.8cm) {$\phantom{ }$};
\end{tikzpicture}
\end{minipage}
\end{align}
\vspace{-1.2cm} \\
where the thick horizontal lines represent the boundary line $\sigma = 0$\,, and the thin curved lines represent the propagator in \eqref{eq:prop}. In particular, a propagator connecting two boundary vertices is given in \eqref{eq:propGab}. All the boundary vertices reside on the $\sigma = 0$ boundary line in the diagrams. Similarly to the case of relativistic string theory in \cite{Abouelsaood:1986gd}, the above diagrams in the sum are only nonzero when $n$ is even. 
A careful analysis shows that the sums of the Feynman diagrams in \eqref{eq:Fdiag1} give
\begin{align} \label{eq:GammaF}
    \sum_{n=0}^\infty \Gamma^\mathscr{F}_n = - \frac{i}{2} \, \Delta (0) \int d\tau \, J^{rI} (\tau) \, \p_\tau t_I{}^1 (\tau) + \text{finite}\,,
\end{align}
and
\begin{align} \label{eq:GammaM}
    \sum_{n=0}^\infty \Gamma^\CM_n = - \frac{1}{2} \, \Delta (0) \int d\tau \, J^{IJ} (\tau) \, \CM_{IJ} (\tau) + \text{finite}\,.
\end{align}
With the definition $\CA = \{ r, I \}$\,, we have
\be
    J^{\CA\CB} = 
    \begin{pmatrix}
        J^{rr} & \quad J^{rJ} \\[2pt]
        J^{Ir} & J^{IJ}
    \end{pmatrix}
\ee
which is defined to be the inverse of
\be
    J_{\CA\CB}
    = G_{\!\CA\CB} - \mathscr{F}_{\!\CA\CC} \, G^{\CC\CD} \mathscr{F}_{\CD\CB}\,,
\ee
where
\be
    G_{\!\CA\CB} = 
    \begin{pmatrix}
        0 & \quad 1 & \quad 0 \\[2pt]
        1 & \quad 0 & \quad 0 \\[2pt]
        0 & \quad 0 & \quad \delta^{A'B'}
    \end{pmatrix},
        \qquad%
    \mathscr{F}_{\!\CA\CB} =
    \begin{pmatrix}
        0 & \quad t_J{}^1 \\[2pt]
        - t_I{}^1 & \quad \mathscr{F}_{IJ}
    \end{pmatrix},
\ee
and $G^{\CA\CB}$ is the inverse of $G_{\CA\CB}$\,. 
In our calculation that leads to \eqref{eq:GammaF}, the total derivatives in \eqref{eq:GammaF} are dropped. This is consistent with that the one-loop effective action is defined up to total derivative terms. However, the full expression of $J^{rI}$ in \eqref{eq:GammaF} is needed for maintaining the gauge invariance when the beta-functions are concerned. See related discussions for the closed string beta-functions in \cite{Yan:2019xsf}. 

Similarly, the sum of the Feynmann diagrams in \eqref{eq:Fdiag2} gives
\be \label{eq:GammaW}
    \sum_{n=0}^\infty \Gamma^\CW_n = - \frac{1}{2} \, \Delta (0) \int d\tau \, J^{I\CA} (\tau) \, \CW_{IJ} (\tau, 0) \, \mathscr{F}^J{}_{\CA} (\tau) + \text{finite}, 
\ee
where, by definition, $\mathscr{F}^J{}_{\CB} = G^{J \CA} \mathscr{F}_{\!\CA\CB}$\,. We have taken a Taylor expansion of $\CW_{IJ} (\tau, \sigma)$ around $\sigma = 0$ and only kept the zeroth order term $\CW_{IJ} (\tau, 0)$ in this calculation.  

Summing \eqref{eq:GammaF}, \eqref{eq:GammaM} and \eqref{eq:GammaW}, and using \eqref{eq:tauraisemut} to convert the indices $0$ and $A'$ back to $i$\,, we find the one-loop boundary effective action,
\begin{align} \label{eq:S1loop}
    S^\text{bdry}_\text{1-loop}
    & = - \frac{1}{2} \, \Delta (0) \! \int_{\p\Sigma} d\tau \Bigl[ i J^{ri} \nabla_{\!\tau} \, t_i{}^1 + J^{ij} M_{ij} - \tfrac{1}{2} \, J^{ia} \mathscr{F}_a{}^j \, \p_i f^\mu \, \p_j f^\nu \mathscr{H}_{\mu\nu\rho} \, \p_\sigma X^\rho_0 \Bigr]\notag \\[2pt]
    & \quad - \frac{i}{2} \, \Delta (0) \!\! \int_{\p\Sigma} d\tau \Bigl[ J^{A'I} \bigl( \Omega_\tau{}^{0A'} \!\! - \! \omega_{A'} \bigr) \, t_I{}^1 - J^{A'B'} \bigl( E^\mu{}_{\!A'} \, \Omega_\mu{}^{0B'} \!\! - \omega_{A'B'} \bigr) \, t_{\tau}{}^1 \Bigr]\,, 
\end{align}
where $\mathscr{H}_{\mu\nu\rho}$ is evaluated at $\sigma = 0$ and $M_{ij}$ is given in \eqref{eq:S(2)}; $J^{ab}$ with  $a = (r, i)$ is defined to be the inverse of
\be \label{eq:defJ}
    J_{ab} = G_{ab} - \mathscr{F}_{ac} \, \mathscr{F}^c{}_b\,.
\ee
 Here, 
\be \label{eq:GF}
    G_{ab} = 
    \begin{pmatrix}
        0 & \quad t_j{}^0 \\[2pt]
        t_i{}^0 & \quad e_{ij}
    \end{pmatrix}\,,
        \qquad%
    \mathscr{F}_{ab} =
    \begin{pmatrix}
        0 & \quad t_j{}^1 \\[2pt]
        - t_i{}^1 & \quad \mathscr{F}_{ij}
    \end{pmatrix},
\ee
 and we  defined $e_{ij} = e_i{}^{A'} e_j{}^{B'}$, with $e_i{}^{A'}$ defined in \eqref{eq:defetF}.
The lower indices can be raised by $G^{ab}$, which is the inverse of $G_{ab}$\,,
\be
    G^{ab} = 
    \begin{pmatrix}
        0 & \quad t^j{}_0 \\[2pt]
        t^i{}_0 & \quad e^{ij}
    \end{pmatrix},
\ee
where $e^{ij} = e^i{}_{A'} e^j{}_{B'}$, with $e^i{}_{A'}$ defined by \eqref{eq:etinv}. The gauge covariance of $S_\text{1-loop}$ in \eqref{eq:S1loop} requires in \eqref{eq:rhozeta}
\be
    \omega_{A'} = \Omega_\tau{}^{0A'},
        \qquad
    \omega_{A'B'} = E^\mu{}_{\!A'} \, \Omega_\mu{}^{0B'}. 
\ee

Applying the boundary conditions \eqref{eq:bdrycondun}, we rewrite \eqref{eq:S1loop} as
\begin{align}
    S_\text{1-loop} = - \frac{1}{2\alpha'} \, \Delta(0) \int_{\p\Sigma} d\tau \,  \ls \tfrac{1}{2} (\lambda - \overline{\lambda}) \, \beta (N) + i \, \p_\tau Y^k \, \beta_k (A) \rs,
\end{align}
where
\begin{subequations} \label{eq:betaNA}
\begin{align}
    \beta(N) & = \alpha' \lr J^{ij} \, K^\mu{}_{ij} \, \tau_\mu{}^1 + \tfrac{1}{2} J^{ia} \mathscr{F}_a{}^j \, \p_i f^\mu \, \p_j f^\nu \, \mathscr{H}_{\mu\nu\rho} \, E^\rho{}_{\!A'} \, t_{A'}{}^1 \rr  + O(\alpha')^2 , \\[6pt]
    \beta_k(A) & = \alpha' \Bigl\{ J^{ri} \, \nabla_{\!k} \, t_i{}^1 - \tfrac{1}{2} \, J^{ia} \, \mathscr{F}_a{}^j \, \p_i f^\mu \, \p_j f^\nu \mathscr{H}_{\mu\nu\rho} \lr \epsilon^A{}_B \, \tau^\rho{}_{\!A} \, t_k{}^B - E^\rho{}_{\!A'} \mathscr{F}_{A'k} \rr \notag \\[2pt]
    & \hspace{2cm} + J^{ij} \Bigr[ \nabla_{i} \mathscr{F}_{jk} + \lr K^\mu{}_{ij} \, \p_k f^\nu + K^\nu{}_{ki} \, \p_{j} f^\mu \rr \mathscr{B}_{\mu\nu} \Bigr] \Bigr\}  + O(\alpha')^2 
\end{align}
\end{subequations}
are the beta-functions for the couplings $N$ and $A_k$ in \eqref{eq:unbrokenra}. 

\subsection{Contribution from the Dilaton} \label{sec:dilaton}

Finally, we discuss the contribution from including a dilaton field in the sigma model action, which is only at the classical level when the lowest order of $\alpha'$ is concerned \cite{Leigh:1989jq}. On a curved worldsheet equipped with a metric $h_{\alpha\beta}$\,, $\alpha, \beta = 1, 2$, we have the dilaton term in the action,
\be
    S_\Phi = \frac{1}{4\pi} \int d^2 \sigma \sqrt{h} \, R[h] \, \Phi [x]\,,
\ee
where $R[h]$ is the worldsheet Ricci scalar and $\Phi [x]$ is the dilaton field. This dilaton term vanishes identically in the flat worldsheet limit, but it contributes nontrivially to the boundary stress energy tensor
\begin{align}
    T_\Phi & = - \p_\sigma \Phi \, \big|_{\sigma = 0} \notag \\[2pt]
    & = \tfrac{1}{2} (\lambda - \overline{\lambda}) \, t_{\!A'}{}^1 \, E^\mu{}_{\!A'} \, \p_\mu \Phi - i \, \p_\tau Y^k \, \p_\mu \Phi \lr \epsilon^A{}_B \, \tau^\mu{}_{\!A} \, t_k{}^B - E^\mu{}_{\!A'} \mathscr{F}_{A'k} \rr. 
\end{align}
The trace of the boundary stress energy tensor is related to the boundary beta-functions $\beta(N)$ and $\beta_k(A)$ by
\be
    T = - \frac{1}{\alpha'} \! \ls \tfrac{1}{2} (\lambda - \overline{\lambda}) \, \beta(N) + i \, \p_\tau Y^k \beta_k (A) \rs .
\ee
Therefore, the beta-functions in \eqref{eq:betaNA} are modified to be
\begin{subequations} \label{eq:betafunctionsPhi}
\begin{align}
    \beta(N) & = \alpha' \lr J^{ij} \, K^\mu{}_{ij} \, \tau_\mu{}^1 + \Theta_\rho \, E^\rho{}_{\!A'} \, t_{A'}{}^1 \rr  + O(\alpha')^2, \\[6pt]
    \beta_k(A) & = \alpha' \Bigl\{ J^{ri} \, \nabla_{\!k} \, t_i{}^1 + J^{ij} \Bigr[ \nabla_{\!i} \mathscr{F}_{jk} + \lr K^\mu{}_{ij} \, \p_k f^\nu + K^\nu{}_{ki} \, \p_{j} f^\mu \rr \mathscr{B}_{\mu\nu} \Bigr] \notag \\[2pt]
    & \hspace{2.75cm} - \Theta_\rho \lr \epsilon^A{}_B \, \tau^\rho{}_{\!A} \, t_k{}^B - E^\rho{}_{\!A'} \mathscr{F}_{A'k} \rr \Bigr\}  + O(\alpha')^2,
\end{align}
\end{subequations}
where
\be \label{eq:Theta}
    \Theta_\rho = \tfrac{1}{2} J^{ia} \mathscr{F}_a{}^j \, \p_i f^\mu \, \p_j f^\nu \mathscr{H}_{\mu\nu\rho}  - \p_\rho \Phi\,.
\ee
Setting the beta-functions $\beta(N)$ and $\beta_k (A)$ to zero gives rise to the equations of motion that determine the backgrounds on which nonrelativistic open string theory can  consistently propagate. Note that the beta-functions of the closed string background fields are not affected at this lowest order in $\alpha'$.

\section{Nonrelativistic Dirac-Born-Infeld Effective Theory} \label{sec:NRDBI}

In this section, we introduce a DBI-like  action that is invariant under the Bargmann symmetry, which describes the effective field theory living on the D-brane. We will show that the classical equations of this DBI action are equivalent to the vanishing one-loop beta-functions in \eqref{eq:betafunctionsPhi}. 

\subsection{Galilean DBI Action from a Nonrelativistic Limit}

To determine the brane action, we start with the relativistic worldvolume DBI action on a D$(d-2)$-brane,
\be \label{eq:relDBI0}
    \widehat{S}_\text{brane} = {T}_{d-2} \int d^{d-1} Y \, e^{-\widehat{\Phi}} \sqrt{- \det \lr \widehat{g}_{ij} + \widehat{\mathcal{F}}_{ij} \rr}\,,
\ee
where $\widehat{g}_{ij} \! = \! \p_i f^\mu \, \p_j f^\nu \, \widehat{G}_{\mu\nu}$\,, $\widehat{\CF}_{ij} \! = \! \p_i f^\mu \, \p_j f^\nu \, \widehat{B}_{\mu\nu} + F_{ij}$\,.
Here, $f^\mu$ is the embedding function that describes how the D-brane is embedded in a $d$-dimensional spacetime.
Consider the following expansions with respect to a large parameter $c$ \cite{Bergshoeff:2019pij}:\footnote{Note that this is different from the D$p$-brane limit considered in \cite{Gomis:2000bd, Brugues:2004an, Brugues:2006yd, Roychowdhury:2019qmp, Pereniguez:2019eoq}, which involves the RR charges. In the D$p$-brane limit, there are no light strings left.} 
\be
    \widehat{G}_{\mu\nu} = c^2 \, \tau_{\mu\nu} + H_{\mu\nu}\,,
        \qquad
    \widehat{B}_{\mu\nu} = - c^2 \, \tau_\mu{}^A \, \tau_\nu{}^B \, \epsilon_{AB} + B_{\mu\nu}\,,
        \qquad
    \widehat{\Phi} = \Phi + \ln |c|\,.
\ee
Then, \eqref{eq:relDBI0} becomes
\be
    \widehat{S}_\text{brane} = {T}_{d-2} \int d^{d-1} Y \, e^{-\Phi} \sqrt{- c^{-2} \, \det \lr h_{ij} + \CF_{ij} - c^2 \, \overline{t}_i \, t_j \rr}\,,
\ee
where $t_i{}^A = \tau_\mu{}^A \, \p_i f^\mu$\,, $h_{ij} = \p_i f^\mu \, \p_j f^\nu \, H_{\mu\nu}$\,, and $\mathcal{F}_{ij} = \p_i f^\mu \, \p_j f^\nu \, B_{ij} + F_{ij}$\,. We also defined $t_i = t_i{}^0 + t_i{}^1$ and $\overline{t}_i = t_i{}^0 - t_i{}^1$\,. Using the identities\footnote{These expressions are non-singular in the limit $m_i{}^A \rightarrow 0$ and $\CF_{ij} \rightarrow 0$ \cite{Bergshoeff:2019pij}.} 
\begin{subequations}
\begin{align}
    \det \! \lr O_{ij} - c^2 \, \overline{t}_i \, t_j \rr & = \lr 1 - c^2 \, t_m \, O^{mn} \, \overline{t}_n \rr \det O_{k\ell}\,, \\[2pt]
    \det \!
    \begin{pmatrix}
        0 & \quad t_j \\[2pt]
        \overline{t}_i & \quad O_{ij}
    \end{pmatrix} & = \lr - t_m \, O^{mn} \, \overline{t}_n \rr \det O_{k\ell}\,, 
\end{align}
\end{subequations}
with $O^{ij}$ the inverse of $O_{ij} \equiv h_{ij} + \CF_{ij}$\,,
we find
\be
    \lim_{c \rightarrow \infty} \widehat{S}_\text{brane} = S_\text{brane}\,,
\ee
where\footnote{A related worldvolume action for D-branes has been considered in \cite{Kluson:2019avy, Kluson:2020aoq}, where the embedding spacetime is taken to be torsional Newton-Cartan spacetime extended with a periodic space direction \cite{Harmark:2017rpg, Harmark:2018cdl}.}
\be \label{eq:Sbrane0}
    S_\text{brane} = T_{d-2} \int d^{d-1} Y \, e^{-\Phi} \sqrt{- \det 
    \begin{pmatrix}
        0 & \quad\! t_j \\[2pt]
        \overline{t}_i & \quad\! h_{ij} + \CF_{ij}
    \end{pmatrix}}\,.
\ee
The action \eqref{eq:Sbrane0} is invariant under the string Newton-Cartan gauge symmetries \eqref{eq:sNCtrnsf}, supplemented with the transformation of $A_i$ in \eqref{eq:deltaAi}. This action is also invariant under the St\"{u}ckelberg transformations \eqref{eq:fredef}, under which the gauge field $A_i$ remains invariant. The invariance under the string Newton-Cartan gauge symmetry and the St\"{u}ckelberg transformations can be shown by using the identity
\be
    \det \begin{pmatrix}
        0 & \quad t_j \\[2pt]
        \overline{t}_i & \quad O_{ij} + a_i \, t_j + b_j \, \overline{t}_i
    \end{pmatrix}
        =
    \det \begin{pmatrix}
        0 & \quad t_j \\[2pt]
        \overline{t}_i & \quad O_{ij}
    \end{pmatrix},
\ee
for arbitrary $O_{ij}$\,, $a_i$ and $b_i$\,.

\subsection{Equations of Motion}

To derive the equations of motion of the brane action \eqref{eq:Sbrane0}, we start with performing a field redefinition using \eqref{eq:fredef} with $C_\mu{}^A = m_\mu{}^A$, and rewrite \eqref{eq:Sbrane0} as
\be \label{eq:Sbrane}
    S_\text{brane} = T_{d-2} \int d^{d-1} Y \, e^{-\Phi} \sqrt{- \det \lr G_{ab} + \mathscr{F}_{ab} \rr}\,,
\ee
with $G_{ab}$ and $\mathscr{F}_{ab}$ defined in \eqref{eq:GF}. 
To extract the appropriate equations of motion from \eqref{eq:Sbrane} to compare with the vanishing beta-functions in \eqref{eq:betafunctionsPhi}, it is useful to use the adapted coordinates introduced in the ``broken phase" in \S\ref{sec:Dsmnos}, where the string Newton-Cartan symmetry is broken to the Bargmann symmetry, with $X^\mu = (y\,, Y^i)$ and 
\be
    f^y = y_0 + N,
        \qquad
    f^i = Y^i.
\ee
Varying the action \eqref{eq:Sbrane} with respect to $N$ and $A_i$\,, we find
\begin{align}
    \delta S_\text{brane} & = - T_{d-2} \int d^{d-1} Y \, e^{- \Phi} \sqrt{-\det\lr G + \mathscr{F} \rr} \notag \\[2pt]
    & \qquad \qquad \qquad \times \Bigl\{ \mathcal{E}_a^{(A)} J^{ak} \delta A_k + \ls \mathcal{E}^{(N)} + \mathcal{E}_a^{(A)} \bigl( J^{ar} + J^{ai} B_{iy} \bigr) \rs \delta N \Bigr\}\,,
\end{align}
where $a = (r, i)$ and
\begin{subequations} \label{eq:EArkN}
\begin{align}
    \mathcal{E}_r^{(A)} & = - J^{ij} \, \nabla_{\!i} \, t_j{}^1 - t_{\!A'}{}^1 \, e^i{}_{\!A'} \, \theta_i\,, \\[2pt]
    \mathcal{E}_k^{(A)} & = J^{ir} \nabla_{\!i} \, t_k{}^1 + J^{ij} \Bigl[ \nabla_{\!i} \mathscr{F}_{jk} + \mathscr{B}_{\mu\nu} \lr K^\mu{}_{ij} \, \p_k f^\nu + K^\nu{}_{ik} \, \p_j f^\mu \rr \Bigr] \!
    + \theta_i \, \mathscr{F}^i{}_k \,, \\[2pt]
    \mathcal{E}^{(N)} & = \tfrac{1}{2} J^{ai} \mathscr{F}_a{}^j \, \p_i f^\mu \, \p_j f^\nu \mathscr{H}_{\mu\nu y} - \p_y \Phi = \Theta_y\,.
\end{align}
\end{subequations}
Here, the definition of $\Theta_y$ matches the one in \eqref{eq:Theta}. We also defined
\be \label{eq:theta}
    \theta_k = \tfrac{1}{2} J^{ai} \mathscr{F}_a{}^j \, \p_i f^\mu \, \p_j f^\nu \, \p_k f^\rho \mathscr{H}_{\mu\nu\rho} -  \, \p_k f^\mu \p_\mu \Phi\,.
\ee
By the definition $t_i{}^1 = \p_i f^\mu \, \tau_\mu{}^1$\,, we have $t_i{}^1 = \tau_i{}^1 + \p_i N$\,. Comparing \eqref{eq:theta} with \eqref{eq:Theta}, it follows that
\be \label{eq:thetaTheta}
    \theta_i = \Theta_i + \Theta_y \, \p_i N.
\ee 
Requiring $\delta S_\text{brane} = 0$ gives
\begin{align} \label{eq:EJAN}
    \mathcal{E}_a^{(A)} J^{ai} = 0\,,
        \qquad
    \mathcal{E}^{(N)} + \mathcal{E}_a^{(A)} J^{ar} = 0\,.
\end{align}
Using the definitions $J^{ac} J_{cb} = \delta^a_b$ and $J_{ab} = G_{ab} - \mathscr{F}_{ac} \mathscr{F}^c{}_b$ in \eqref{eq:defJ}, we find that \eqref{eq:EJAN} is equivalent to
\begin{subequations} \label{eq:ErAEkA}
\begin{align}
    \mathcal{E}_r^{(A)} & = \lr \mathcal{E}_a^{(A)} J^{ai} \rr J_{ir} + \lr \mathcal{E}_a^{(A)} J^{ar} \rr J_{rr} = - \mathcal{E}^{(N)} \, t_{A'}{}^1 \, t_{A'}{}^1, \\[2pt]
    \mathcal{E}_k^{(A)} & = \lr \mathcal{E}_a^{(A)} J^{ai} \rr J_{ik} + \lr \mathcal{E}_a^{(A)} J^{ar} \rr J_{rk} = - \mathcal{E}^{(N)} \lr t_k{}^0 - t_0{}^1 \, t_k{}^1 - t_{A'}{}^1 \mathscr{F}_{A'k} \rr.
\end{align}
\end{subequations}
Plugging \eqref{eq:EArkN} into \eqref{eq:ErAEkA}, we find
\begin{subequations} \label{eq:eomDBI}
\begin{align}
    0 & = J^{ij} \, \nabla_{\!i} \, t_j{}^1 + t_{\!A'}{}^1 \ls e^i{}_{\!A'} \lr \Theta_i + \Theta_y \, \p_i N \rr - t_{\!A'}{}^1 \, \Theta_y \rs, \\[2pt]
    0 & = J^{ir} \nabla_{\!i} \, t_k{}^1 + J^{ij} \Bigl[ \nabla_{\!i} \mathscr{F}_{jk} + \mathscr{B}_{\mu\nu} \lr K^\mu{}_{ij} \, \p_k f^\nu + K^\nu{}_{ik} \, \p_j f^\mu \rr \Bigr] \notag \\[2pt]
    & \hspace{2.05cm} + \lr \Theta_i + \Theta_y \, \p_i N \rr \mathscr{F}^i{}_k + \Theta_y \lr t_k{}^0 - t_0{}^1 \, t_i{}^1 - t_{\!A'}{}^1 \, \mathscr{F}_{\!A'k} \rr.
\end{align}
\end{subequations}

In order to show that the equations of motion in \eqref{eq:eomDBI} match with the vanishing beta-functions in \eqref{eq:betafunctionsPhi}, we need to use the identities in \eqref{eq:pretE} and \eqref{eq:tK}, with
\be \label{eq:id1}
    \tau^i{}_1 = 0\,,
        \quad
    \tau^y{}_ 0 = - \tau^i{}_0 \, \tau_i{}^1,
        \quad
    E^y{}_{A'} = - E^i{}_{A'} \, \tau_i{}^1,
        \quad
    \tau_\mu{}^1 K^\mu{}_{ij} = \nabla_{\!i} \, t_j{}^1 = \nabla_{\!j} \, t_i{}^1.
\ee
Moreover, using the prescriptions in  \eqref{eq:pretE} and \eqref{eq:invsubm}, together with the invertibility condition in \eqref{eq:etinv}, we find
\be \label{eq:invtaui}
    \tau^i{}_0 = t^i{}_0\,,
        \qquad%
    E^i{}_{A'} = e^i{}_{A'}\,.
\ee
Applying \eqref{eq:id1} and \eqref{eq:invtaui} to \eqref{eq:eomDBI}, we find
\begin{subequations}
\begin{align}
    0 & = J^{ij} \, K^\mu{}_{ij} \, \tau_\mu{}^1 + t_{\!A'}{}^1 \, E^\rho{}_{\!A'} \, \Theta_\rho\,, \\[2pt]
    0 & = J^{ir} \nabla_{\!k} \, t_i{}^1 + J^{ij} \Bigl[ \nabla_{\!i} \mathscr{F}_{jk} + \mathscr{B}_{\mu\nu} \lr K^\mu{}_{ij} \, \p_k f^\nu + K^\nu{}_{ik} \, \p_j f^\mu \rr \Bigr] \notag \\[2pt]
    & \hspace{2.1cm} - \Theta_\rho \lr \epsilon^A{}_B \, \tau^\rho{}_{\!A} \, t_k{}^B - E^\rho{}_{\!A'} \mathscr{F}_{A'k} \rr.
\end{align}
\end{subequations}
These are precisely the same equations from setting the beta-functions in \eqref{eq:betafunctionsPhi} to zero.

\subsection{Galilean Electrodynamics on a Newton-Cartan Background}

It is interesting to consider the worldvolume action \eqref{eq:Sbrane0} in a simple case where we assume that $y$ is an isometry direction and take the following specifications on the boundary:
\be
    \tau_i{}^1 = m_i{}^1 = m_y{}^1 = m_y{}^0 = B_{\mu\nu} = \Phi = 0\,.
\ee
We therefore consider a zero B-field and dilaton background.
Then, the worldvolume action \eqref{eq:Sbrane0} reduces to 
\begin{align} \label{eq:Sbranesp}
    S_\text{brane} & = T_{d-2} \int d^{d-1} Y \, \sqrt{- \det (\mathscr{G}_{ab} + \mathcal{F}_{ab})}\,,
\end{align}
where
\be
    \mathscr{G}_{ab} = 
    \begin{pmatrix}
        0 & \,\, \tau_j{}^0 \\[2pt]
        \tau_i{}^0 & \,\, H_{ij}
    \end{pmatrix},
        \qquad
    \mathcal{F}_{ab} = \begin{pmatrix}
        0 & \,\, \p_j N \\[2pt]
        - \p_i N & \,\, {F}_{ij}
    \end{pmatrix}, 
\ee
and $H_{ij} = E_{ij} - \bigl( m_i{}^0 \, \tau_j{}^0 + m_j{}^0 \, \tau_i{}^0 \bigr)$\,, with $E_{ij} = E_i{}^{A'} E_j{}^{B'}$.
The action in \eqref{eq:Sbranesp} can be further rewritten as
\begin{align} \label{eq:SbranescrGF}
    S_\text{brane} & = T_{d-2} \int d^{d-1} Y \sqrt{-\mathscr{G}} \, \sqrt{\det \! \lr \delta^a_b +
        \mathscr{G}^{ac} \mathcal{F}_{cb} \rr}\,,
\end{align}
where $\mathscr{G}$ is the determinant of $\mathscr{G}_{ab}$ and $\mathscr{G}^{ab}$ is the inverse of $\mathscr{G}_{ab}$\,, with
\begin{align}
    \mathscr{G} & = - \bigl( \tau_k{}^0 H^{k\ell} \tau_\ell{}^0 \bigr) \, \det H_{ij}\,, 
        \qquad
    \mathscr{G}^{ab} =
    \begin{pmatrix}
        2 \, \phi & \,\, T^j{}_0 \\[2pt]
        T^i{}_0 & \,\, E^{ij}
    \end{pmatrix},
\end{align}
where $E^{ij} = E^i{}_{A'} E^j{}_{B'}$. Note that $\mathscr{G}$ is independent of $m_\mu{}^0$.
We also defined
\be
    \phi = m_i{}^0 \, \tau^i{}_0 + \frac{1}{2} \, m_i{}^0 \, E^{ij} \, m_j{}^0, 
        \qquad
    T^i = \tau^i{}_0 + E^{ik} \, m_k{}^0.
\ee
Here, $\phi$ is related to the Newton potential. At the quadratic order in fields, \eqref{eq:SbranescrGF} gives Galilean Electrodynamics on a Newton-Cartan background,
\begin{align} \label{eq:GEDcurved}
    & \quad S_\text{GED} = \frac{1}{4 g^2} \int d^{25} Y \sqrt{-\mathscr{G}} \, \mathscr{G}^{ab} \mathcal{F}_{bc} \, \mathscr{G}^{cd} \mathcal{F}_{da} \notag \\[2pt]
        & = \frac{1}{g^2} \! \int \! d^{25} Y \!\sqrt{-\mathscr{G}} \Bigl[ \tfrac{1}{2} \bigl( \,  T^{i} \, T^j \! - \! 2 \, \phi \, E^{ij} \bigr) \nabla_{\!i} N \, \nabla_{\!j} N + E^{ik} \! \lr T^\ell \nabla_{\!i} N \! - \! \tfrac{1}{4} E^{j\ell} F_{ij} \rr F_{k\ell} \Bigr]. 
\end{align}

Alternatively, we can also start with the equivalent worldvolume action \eqref{eq:Sbrane}, which now reduces to
\be \label{eq:Sbranelast}
    S_\text{brane} = T_{d-2} \int d^{d-1} Y \, \sqrt{- \det \lr G_{ab} + \mathscr{F}_{ab} \rr}\,,
\ee
where
\be
    G_{ab} = 
    \begin{pmatrix}
        0 & \,\, \tau_j{}^0 \\[2pt]
        \tau_i{}^0 & \,\, E_{ij}
    \end{pmatrix},
        \qquad
    \mathscr{F}_{ab} = \begin{pmatrix}
        0 & \,\, \p_j N\, \\[2pt]
        - \p_i N & \,\, \mathscr{F}_{ij}\,
    \end{pmatrix}, 
\ee
and $\mathscr{F}_{ij} = F_{ij} + m_i \, \p_{j} N - m_j \, \p_{i} N$, or, equivalently,
\be
\mathscr{F}_{ij} = \p_i \mathscr{A}_j - \p_j \mathscr{A}_i + N \lr \p_i m_j{}^0 - \p_j m_i{}^0 \rr,    
        \qquad
    \mathscr{A}_i = A_i - m_i{}^0 \, N\,.
\ee
At the quadratic order in fields, \eqref{eq:Sbranelast} gives an alternative form of Galilean Electrodynamics on a Newton-Cartan background,
\begin{align} \label{eq:GEDcurved2}
    S_\text{GED} & = \frac{1}{4 g^2} \int d^{25} Y \sqrt{-G} \, G^{ab} \mathscr{F}_{bc} \, G^{cd} \mathscr{F}_{da} \notag \\[2pt]
        & = \frac{1}{g^2} \int d^{25} Y \sqrt{-G} \Bigl[ \tfrac{1}{2} \tau^{i}{}_0 \, \tau^j{}_0 \nabla_{\!i} N \, \nabla_{\!j} N + E^{ik} \! \lr \tau^\ell{}_0 \nabla_{\!i} N - \tfrac{1}{4} E^{j\ell} \mathscr{F}_{ij} \rr \mathscr{F}_{k\ell} \Bigr],
\end{align}
which is equivalent to \eqref{eq:GEDcurved}. Here, $G$ is the determinant of $G_{ab}$ and $G = \mathscr{G}$.

The equivalent actions \eqref{eq:GEDcurved} and \eqref{eq:GEDcurved2} coincide with the actions considered in \cite{Festuccia:2016caf}.
In the flat spacetime limit, with $\tau_i{}^0 = \delta_i^0$\,, $E_i{}^{A'} = \delta_i^{A'}$ and $m_i{}^0 = 0$\,, both \eqref{eq:GEDcurved} and \eqref{eq:GEDcurved2} reduce to the action of Galilean Electrodynamics in \eqref{eq:GEDflat}.

\section{Conclusions} \label{sec:conclusions}

In this paper, 
we have studied the Dirichlet nonlinear sigma model that describes nonrelativistic open strings ending on   D-branes in the string Newton-Cartan geometry, Kalb-Ramond and dilaton closed string background. Having a nonrelativistic open string spectrum  requires the D-branes to be transverse to the longitudinal spatial direction in the string Newton-Cartan geometry background. We have computed  the beta-functions for the open string vertex operators on a single D$(d-2)$-brane, transverse to the longitudinal spatial direction. Self-consistency of nonrelativistic open string theory requires setting these beta-functions to zero, which gives rise to the equations of motion that govern the dynamics of the D-brane. We also showed that the same set of equations of motion arise from an action principle in \eqref{eq:Sbrane0}, which is the worldvolume nonrelativistic DBI action of the D-brane. In a companion paper \cite{Tdualopenstring}, we will consider T-duals of nonrelativistic open string theory.

At  leading  order in $\alpha'$ and in the flat closed string background, the DBI action in \eqref{eq:Sbrane0} gives rise to Galilean Electrodynamics (GED). GED is a non-dynamical $U(1)$ gauge theory that is invariant under a Galilean boost transformation, and
has been studied at the classical level in \cite{Santos:2004pq, Bergshoeff:2015sic, Festuccia:2016caf, Banerjee:2019axy}. In \cite{scalarGED}, the one-loop beta-functions of Galilean electrodynamics coupled to a Schr\"{o}dinger scalar in $2+1$ dimensions are computed, where the renormalization of the dynamical Schr\"{o}dinger scalar receives highly nontrivial contributions from interactions with the non-dynamical gauge sector. There is an extra scalar in addition to the $U(1)$ gauge field in GED, which finds a natural interpretation in nonrelativistic open string theory as the Nambu-Goldstone boson from spontaneously breaking the string Newton-Cartan symmetry algebra to the Bargmann symmetry algebra.

We also considered the low energy effective action  on $n$ coincident D-branes, where our worldsheet analysis has led to a novel nonrelativistic $U(n)$ Yang-Mills theory.
 Finally,   we have shown how to incorporate open  string winding modes and demonstrated that in spite of the nonlocal features of  wound open strings that  the   nontrivial dynamics and spectrum of open strings is neatly captured by a   gauge theory in one higher dimension, and interpreted the additional dimension as conjugate to winding number. 

\acknowledgments

We would like to thank the organizers and participants at the online seminar series on ``Non-Lorentzian Geometries: Non-Relativistic String Theory" (June 2020) for stimulating discussions, where results in this paper were first presented. 
This research is supported in part by Perimeter Institute for Theoretical Physics.
Research at Perimeter Institute is supported in part by the Government of Canada through the Department of Innovation, Science and Economic Development Canada and by the Province of Ontario through the Ministry of Colleges and Universities.

\newpage

\bibliographystyle{JHEP}
\bibliography{nros}

\end{document}